\documentclass[11pt,a4paper,aps,tightenlines,
preprintnumbers,nofootinbib,showkeys,superscriptaddress]{revtex4-1}
\pdfoutput=1
\usepackage{fullpage}
\usepackage{float}
\usepackage{amsfonts}
\usepackage{amsmath}
\usepackage{slashed}
\usepackage{mathrsfs}
\usepackage[scr=rsfs,cal=boondox]{mathalfa}
\usepackage{amssymb}
\usepackage{graphicx}
\usepackage{makeidx}
\usepackage{cancel}
\usepackage{eepic}
\usepackage{epsfig}
\usepackage{latexsym}
\usepackage{mathtools}
\usepackage[dvipsnames]{xcolor}
\usepackage{float}
\usepackage{multirow}
\usepackage[export]{adjustbox}
\usepackage{xurl,hyperref}
\usepackage{enumitem}
\hypersetup{colorlinks=true,citecolor=red,linkcolor=NavyBlue,urlcolor=NavyBlue}
\usepackage[utf8]{inputenc}
\usepackage[caption=false]{subfig}
 
\usepackage{natbib}
\usepackage{mathptmx}

\def\ba{\begin{array}}       
\def\ea{\end{array}}

\def\beq{\begin{eqnarray}}
\def\eeq{\end{eqnarray}}

\begin{document}
\title{Two-Loop Contributions to the Anomalous Chromomagnetic Dipole Moment of the Top Quark in Two-Higgs-Doublet Models}
\author{Subhadip Bisal}
\email{subhadip.b@iopb.res.in}
\affiliation{Institute of Physics, Sachivalaya Marg, Bhubaneswar 751 005, India}
\affiliation{Homi Bhabha National Institute, Training School Complex, Anushakti Nagar, Mumbai 400 094, India}

\date{\today}
\hspace{1.0cm}\preprint{IOP/BBSR/2024-02}

\begin{abstract}
\noindent
\textbf{Abstract:} 
 We present the analytical results for the anomalous chromomagnetic moment of a quark at the one-loop and two-loop levels through model-independent parameterizations, considering the external gluon to be off-shell. Then, we consider different types of two-Higgs doublet models with the momentum transfer of the external gluon, $q^2 = \pm M_Z^2$, and $0$, for the numerical evaluations thereof for the top quark. We find that the contributions in the two-Higgs-doublet models are on the order of $\mathcal{O}(10^{-3})$ at the one-loop level and $\mathcal{O}(10^{-4})$ at the two-loop level, where the one-loop and two-loop contributions interfere destructively, resulting in the overall contributions being on the order of $\mathcal{O}(10^{-3})$ with a positive sign.
\end{abstract}

\maketitle

\section{Introduction}
The planned next generation of high-energy colliders will test the Standard Model (SM) with high precision and explore higher energies in the search for new physics (NP).
The NP may manifest itself in two ways: either through direct signals entailing the emergence of new particles or via deviations from the SM predictions concerning known particles. In certain instances, indirect effects can provide evidence of new physics beyond the SM (BSM), even prior to the discovery of new particles. Consequently, examining corrections of new physics to observables is essential, too. 
Amongst these observables, the anomalous magnetic and electric dipole moments are separate cornerstones within the realm of BSM physics. Just as leptons have a magnetic moment, quarks are also associated with a magnetic moment known as the anomalous chromomagnetic dipole moment (CMDM), which first appears at the one-loop level when there is an external gluon emission.

The CMDM of the top quark at the Large Hadron Collider (LHC) has been recently measured by the CMS collaboration, using $pp$ collisions at a center-of-mass energy of 13 TeV and an integrated luminosity of 35.9 ${\rm fb^{-1}}$~\cite{CMS:2019kzp}. According to their report,
\begin{align}
    \hat{\mu}_t^{\rm Exp} = -0.024^{+0.013}_{-0.009}({\rm stat})^{+0.016}_{-0.011}({\rm syst})~,
\end{align}
whereas for the chromoelectric dipole moment (CEDM), $|\hat{d}_t^{\rm Exp}|<0.03$ at $95\%$ C.L. 
The anomalous CMDM, $\hat{\mu}_q$, in the SM, is generated at the one-loop level, and the contributions come from quantum chromodynamics (QCD), electroweak (EW), and Yukawa sectors~\cite{Choudhury:2014lna, Aranda:2018zis, Aranda:2020tox, Hernandez-Juarez:2020drn}. An important characteristic of this property refers to the presence of an infrared (IR) divergence originating from the one-loop Feynman diagram associated with the QCD non-Abelian triple gluon vertex. This issue arises when the external gluon is taken on-shell, i.e., $q^2=0$, where $q$ is the momentum transfer of the external gluon, as previously noted in Refs.~\cite{Choudhury:2014lna, Bermudez:2017bpx, Aranda:2018zis}. Ref.~\cite{Aranda:2020tox} also studied this issue using dimensional regularization (DR), where they identified the IR divergence by a $1/\epsilon_{\rm IR}$ pole. They showed that the IR divergence comes from the two-point Passarino-Veltman (PV) function $\mathbf{B}_0(q^2,0,0)$ when $q^2=0$. Ref.~\cite{Choudhury:2014lna, Aranda:2020tox} proposed to evaluate the CMDM at a large momentum transfer $q^2=\pm M_Z^2$. Justification for this choice arises from the characterization of the strong running coupling constant at that conventional scale in perturbative QCD (pQCD), $\alpha_s(M_Z^2)=0.1179$~\cite{ParticleDataGroup:2020ssz}.

Since the CMDM could significantly receive contributions from BSM physics, numerous computations have been documented in the literature within the extension of the SM, such as the two-Higgs-doublet model (THDM)~\cite{Gaitan:2015aia}, the four-generation THDM~\cite{Hernandez-Juarez:2018uow, Martinez:2001qs, Martinez:2007qf}, models with a heavy $Z^\prime$ gauge boson~\cite{Aranda:2018zis}, little Higgs models~\cite{Cao:2008qd, Ding:2008nh, Cisneros-Perez:2024onx},  the minimal
supersymmetric standard model (MSSM)~\cite{Aboubrahim:2015zpa}, unparticle models~\cite{Martinez:2008hm}, vectorlike multiplet models~\cite{Ibrahim:2011im}, reduced 331 model~\cite{Hernandez-Juarez:2020gxp, Martinez:2007qf} etc. All the aforementioned BSM models studied the CMDM at the leading order (LO) or one-loop level.
The CMDM of the top quark within the context of a decoupling effective Lagrangian approach using the radiative decay of $b\to s\gamma$ was studied in Ref.~\cite{Martinez:1996cy}. The next-to-leading order (NLO) QCD corrections to the top quark CMDM in $t\bar{t}$ production is discussed in Ref.~\cite{BuarqueFranzosi:2015jrv}. They found that NLO corrections enhance the contribution from the CMDM by about $50\%$ at the LHC and substantially decrease the renormalization and factorization scale dependence.

In this work, we present a study on the contributions of the off-shell CMDM of the top quark in the aligned THDM (ATHDM)~\cite{Pich:2009sp, Pich:2010ic, Karan:2023kyj, Eberhardt:2020dat} and in the conventional THDMs with $\mathcal{Z}_2$ symmetries~\cite{Eberhardt:2013uba, Chowdhury:2015yja, Cacchio:2016qyh, Eberhardt:2018lub, Chowdhury:2017aav, Eberhardt:2017ulj, Eberhardt:2015ypa, Eberhardt:2014kaa, Gunion:2002zf} in one-loop and two-loop order, which has not been studied in the literature to the best of my knowledge.
THDMs are well-motivated extensions of the SM with five physical Higgs bosons: two CP-even
states $h$ and $H$, one CP-odd state $A$, and two charged states $H^\pm$.
The one-loop contributions to the CMDM are mediated by these neutral and charged Higgs bosons. Apart from these contributions, there are two-loop Barr-Zee type diagrams~\cite{Barr:1990vd} involving a closed fermionic loop. We know that the two-loop Barr-Zee type diagrams provide significant contributions in the context of dipole moment. Here, we present a full two-loop calculation and the analytical results for the Barr-Zee type diagrams arising from the THDMs by considering the external gluon to be off-shell ($q^2\ne0$). Finally, we present the numerical results for the top quark CMDM in the THDMs across time-like ($q^2>0$), space-like ($q^2<0$), and light-like ($q^2=0$) scenarios.

The manuscript is structured as follows: Sec.\ref{sec:CMDMintro} presents a brief introduction to the CMDM and CEDM. In Sec.\ref{sec:oneloop}, the one-loop contributions to CMDM are presented, while Sec.\ref{sec:twoloop} elaborates on the two-loop calculations contributing to the CMDM. In Sec.\ref{sec:modelTHDM}, a discussion on the ATHDM as well as the conventional THDMs with $\mathcal{Z}_2$ symmetries is given. The numerical results are described in Sec.\ref{sec:results}, wherein Sec.\ref{sec:ATHDMcmdm} and Sec.\ref{sec:THDMz2} delineate the outcomes within the ATHDM and the conventional THDMs with $\mathcal{Z}_2$ symmetries, respectively. Finally, the conclusion is provided in Sec.\ref{sec:conclusion}.

\section{The Chromomagnetic Dipole Moment}
\label{sec:CMDMintro}
The interaction between quark-antiquark pairs and gluons, as described by the Lagrangian incorporating the effective chromoelectromagnetic dipole moment (CEMDM), is expressed as~\cite{Bernreuther:2013aga, CMS:2016piu, Haberl:1995ek}:
\begin{align}
    \mathcal{L}_{q\bar{q}g} = -g_s\bar{q}_A\gamma^{\mu}q_B g_{\mu}^a T^a_{AB} + \mathcal{L}_{\rm eff}~,
\end{align}
with 
\begin{align}
    \mathcal{L}_{\rm eff} = -\frac{1}{2}\bar{q}_A \sigma^{\mu\nu} \big(\mu_q + i d_q\gamma^5\big) q_B G^a_{\mu\nu} T^a_{AB}~,
    \label{Eq:effCEMDM}
\end{align}
where $\sigma^{\mu\nu}=\frac{i}{2}\big[\gamma^\mu, \gamma^\nu\big]$, $\mu_q$ is the CP-conserving part, called chromomagnetic form factor, $d_q$ is the CP-violating part, called chromoelectric form factor, $T^a_{AB}$ is the colour generator of $SU(3)_C$ with $A$ and $B$ being the quark colour indices, $G_{\mu\nu}^a = \partial_\mu g_\nu^a- \partial_\nu g_\mu^a -g_sf_{abc}g_\mu^b g_\nu^c$ is the gluon field strength tensor. In the SM, the CMDM can be generated at one-loop level~\cite{Choudhury:2014lna, Bermudez:2017bpx, Martinez:2007qf}, whereas the CEDM first appears at three-loop level~\cite{Czarnecki:1997bu}. Customarily, one can define the CMDM and CEDM in their dimensionless form~\cite{ParticleDataGroup:2020ssz, Bernreuther:2013aga, CMS:2016piu, Haberl:1995ek} as

\begin{align}
    \hat{\mu}_q = \frac{m_q}{g_s} \mu_q~,\hspace{1cm} \hat{d}_q = \frac{m_q}{g_s} d_q~,
    \label{eq:cemdm}
\end{align}
where $m_q$ is the mass of quark, $g_s=\sqrt{4\pi\alpha_s}$ with $\alpha_s$ is the strong coupling constant. In general, CEMDM may contain absorptive parts, which becomes apparent, for instance, when the momentum transfer assumes a timelike nature, i.e., $q^2>0$, particularly evident in processes such as top-antitop ($t\bar{t}$) production through proton-proton ($pp$) collisions~\cite{Bernreuther:2013aga, CMS:2016piu}. From Eq.~\eqref{Eq:effCEMDM}, one can write the effective vertex giving rise to the CEMDM as 
\begin{align}
    \Gamma^\mu = \sigma^{\mu\nu} q_\nu \big(\mu_q+ id_q\gamma^5\big) T^a_{AB}~,
\end{align}
where $q_\nu$ is the gluon momentum transfer. The corresponding invariant amplitude can be written as
\begin{align}
    \mathcal{M} = \mathcal{M}^\mu \epsilon^a_\mu(\Vec{q})~,
\end{align}
where $p$ and $p^\prime$ are the momenta of the external quarks so that $q=p^\prime-p$. The Lorentz structure can be written as 
\begin{align}
    \hspace{1cm}\mathcal{M}^\mu = \bar{u}(p^\prime)\Gamma^\mu u(p)~.
\end{align}
\section{One-loop contributions}
\label{sec:oneloop}
\begin{figure}[H]
	\centering
	\includegraphics[width=0.95\linewidth]{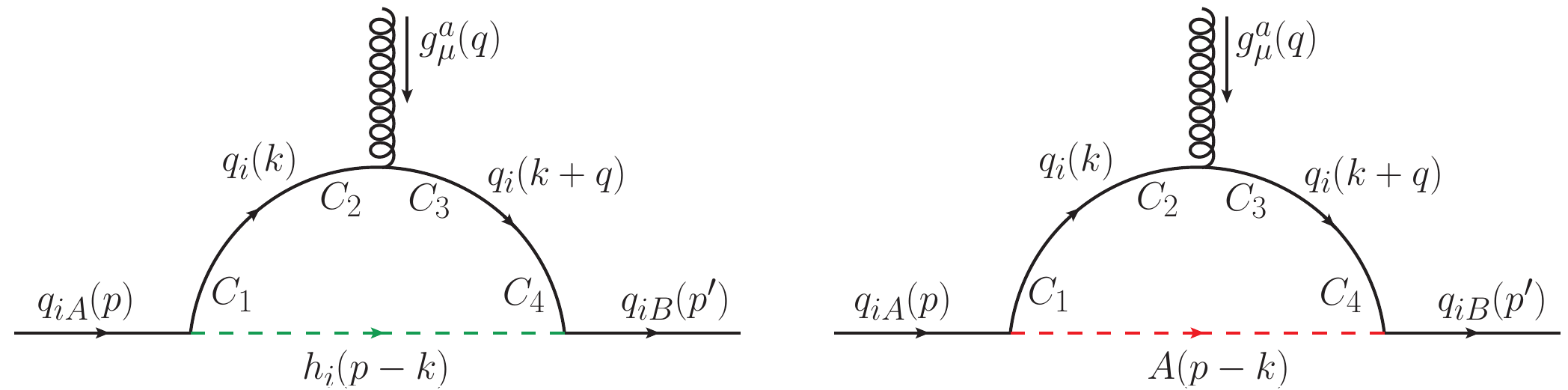}\\
	\,\,(a)\hspace{7.5cm}(b)\\
    \includegraphics[width=0.47\linewidth]{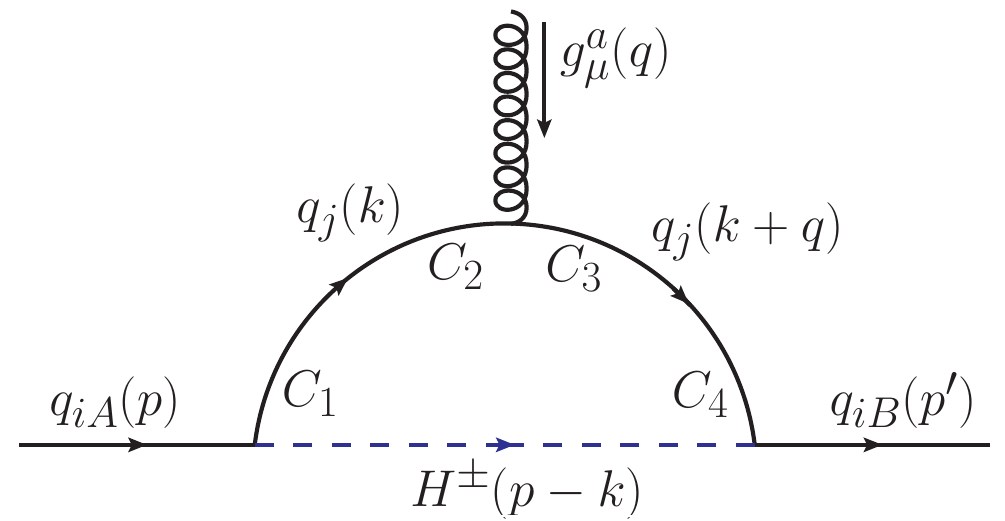}\\
    (c)
	\caption{One-loop diagrams contributing to the CMDM of a quark of flavor $``i"$ in an extended Higgs sector with two CP-even scalars ($h, H$), one pseudoscalar ($A$), and two charged Higgs bosons ($H^\pm$).}
	\label{fig:oneloopcmdm}
\end{figure}

The one-loop diagrams with momentum assignments, which contribute to the CMDM of quarks in an extended Higgs sector with two CP-even Higgs scalars $\big(h_i\in\bigl\{h, H\bigr\}\big)$, one pseudoscalar ($A$), and two charged Higgs bosons ($H^\pm$), are illustrated in Fig.~\ref{fig:oneloopcmdm}.
Here, we present the analytical expressions for the CMDM applicable to any quark of flavor ``$i$" in a model-independent manner. Finally, we discuss the different types of THDMs and their contributions to the CMDM of the top quark.

The amplitude for the diagram depicted in Fig.~\ref{fig:oneloopcmdm}a can be expressed as 
\begin{align}
\mathcal{M}^{\mu}_{(a)\rm 1L}= &\int \frac{d^dk}{(2\pi)^d} \bar{u}(p^\prime)\Bigg[\big(-iy_{q_i q_i h_i}\delta_{C_4B}\big)\Biggl\{\frac{i\big(\slashed{k}+\slashed{q}+m_{q_i}\big)}{\big(k+q\big)^2-m_{q_i}^2}\delta_{C_3C_4}\Biggr\}\big(-ig_s T^a_{C_2C_3}\gamma^\mu\big)\Biggl\{\frac{i\big(\slashed{k}+m_{q_i}\big)}{k^2-m_{q_i}^2}\delta_{C_1C_2}\Biggr\}\nonumber\\
&\times\big(-iy_{q_i q_i h_i}\delta_{AC_1}\big)\Biggl\{\frac{i}{\big(p-k\big)^2-m_{h_i}^2}\Biggr\}\Bigg]u\big(p\big)~,
\label{eq:1_1La}
\end{align}
where $y_{q_i q_i h_i}$'s are the Yukawa couplings of the CP-even Higgs scalars with the quarks, $m_{q_i}$ is the mass of quark of flavor ``$i$", $m_{h_i}$'s are the masses of CP-even Higgs scalars, and $A, B, C_1,..., C_4$ represent the color indices for the quarks.

Here, we use {\sc Package-X}~\cite{Patel:2015tea, Patel:2016fam} to calculate the amplitudes of the diagrams shown in Fig.~\ref{fig:oneloopcmdm}. To calculate the CMDM, first, we take the coefficient of $\sigma^{\mu\nu}q_\nu$ and then use Eq.~\eqref{eq:cemdm}. The CMDM can be extracted from the amplitude in Eq.~\eqref{eq:1_1La} as

\begin{align}
    \hat{\mu}_{q_i}^{(a), \rm 1L} =& -\frac{y_{q_i q_i h_i}^2 m^2_{q_i}}{16\pi^2}\Big[\mathbf{C}_0-\mathbf{C}_{11}\Big]~.
    \label{eq:cp-evenloop}
\end{align}

The amplitude for the diagram in Fig.~\ref{fig:oneloopcmdm}b can be written in a similar way as in Eq.~\eqref{eq:1_1La}, except the fact that the Yukawa coupling for the pseudoscalar Higgs includes a $\gamma^5$-term.
Therefore, the CMDM from Fig.~\ref{fig:oneloopcmdm}b can be obtained as 

\begin{align}
    \hat{\mu}_{q_i}^{(b), \rm 1L} =& -\frac{y_{q_i q_i h_i}^2 m^2_{q_i}}{16\pi^2}\Big[\mathbf{C}_0+2\mathbf{C}_{1} + \mathbf{C}_{11}\Big]~.
    \label{eq:cp-oddloop}
\end{align}

The PV functions for Eq.~\eqref{eq:cp-evenloop} and \eqref{eq:cp-oddloop}, are defined as
\begin{align*}
     \mathbf{C}_k=&\mathbf{C}_k\bigg[m_{q_i}^2, 2\biggl\{\frac{1}{2}\big(2m_{q_i}^2-q^2\big)-m_{q_i}^2\biggr\}+m_{q_i}^2+q^2,q^2; m_{q_i}, m_{h_i/A}, m_{q_i}\bigg]~,\\
    \mathbf{C}_{k\ell}=&\mathbf{C}_{k\ell}\bigg[m_{q_i}^2, 2\biggl\{\frac{1}{2}\big(2m_{q_i}^2-q^2\big)-m_{q_i}^2\biggr\}+m_{q_i}^2+q^2,q^2; m_{q_i}, m_{h_i/A}, m_{q_i}\bigg]~.
\end{align*}

Similarly, the contribution to the CMDM from Fig.~\ref{fig:oneloopcmdm}c is given by

\begin{align}
    \hat{\mu}_{q_i}^{(c), \rm 1L} =& -\frac{y_{q_i q_j H^\pm}^2 m_{q_i} \lvert V_{ij}\rvert^2}{16\pi^2}\Bigg[C_LC_R m_{q_j}\big(\mathbf{C}_0+\mathbf{C}_1\big) - \frac{1}{2}\big(C_L^2+C_R^2\big) m_{q_i}\big(\mathbf{C}_1+ \mathbf{C}_{11}\big) \Bigg]~,
    \label{eq:hpmloop}
\end{align}
where $V_{ij}$ is the element of the CKM matrix, $C_L$, and $C_R$ are coefficients of the left-handed and right-handed projection operators in the $H^{\pm}q_iq_j$ ($i\ne j$) interaction where $q_i$ and $q_j$ are up(down) and down(up) type quarks, respectively. The PV functions for Eq.~\eqref{eq:hpmloop} are defined as 
\begin{align*}
    \mathbf{C}_k=&\mathbf{C}_k\bigg[m_{q_i}^2, 2\biggl\{\frac{1}{2}\big(2m_{q_i}^2-q^2\big)-m_{q_i}^2\biggr\}+m_{q_i}^2+q^2,q^2; m_{q_j}, m_{H^\pm}, m_{q_j}\bigg]~,\\
    \mathbf{C}_{k\ell}=&\mathbf{C}_{k\ell}\bigg[m_{q_i}^2, 2\biggl\{\frac{1}{2}\big(2m_{q_i}^2-q^2\big)-m_{q_i}^2\biggr\}+m_{q_i}^2+q^2,q^2; m_{q_j}, m_{H^\pm}, m_{q_j}\bigg]~.
\end{align*}

\section{Two-loop contributions}
\label{sec:twoloop}

\begin{figure}[H]
	\centering
	\includegraphics[width=0.86\linewidth]{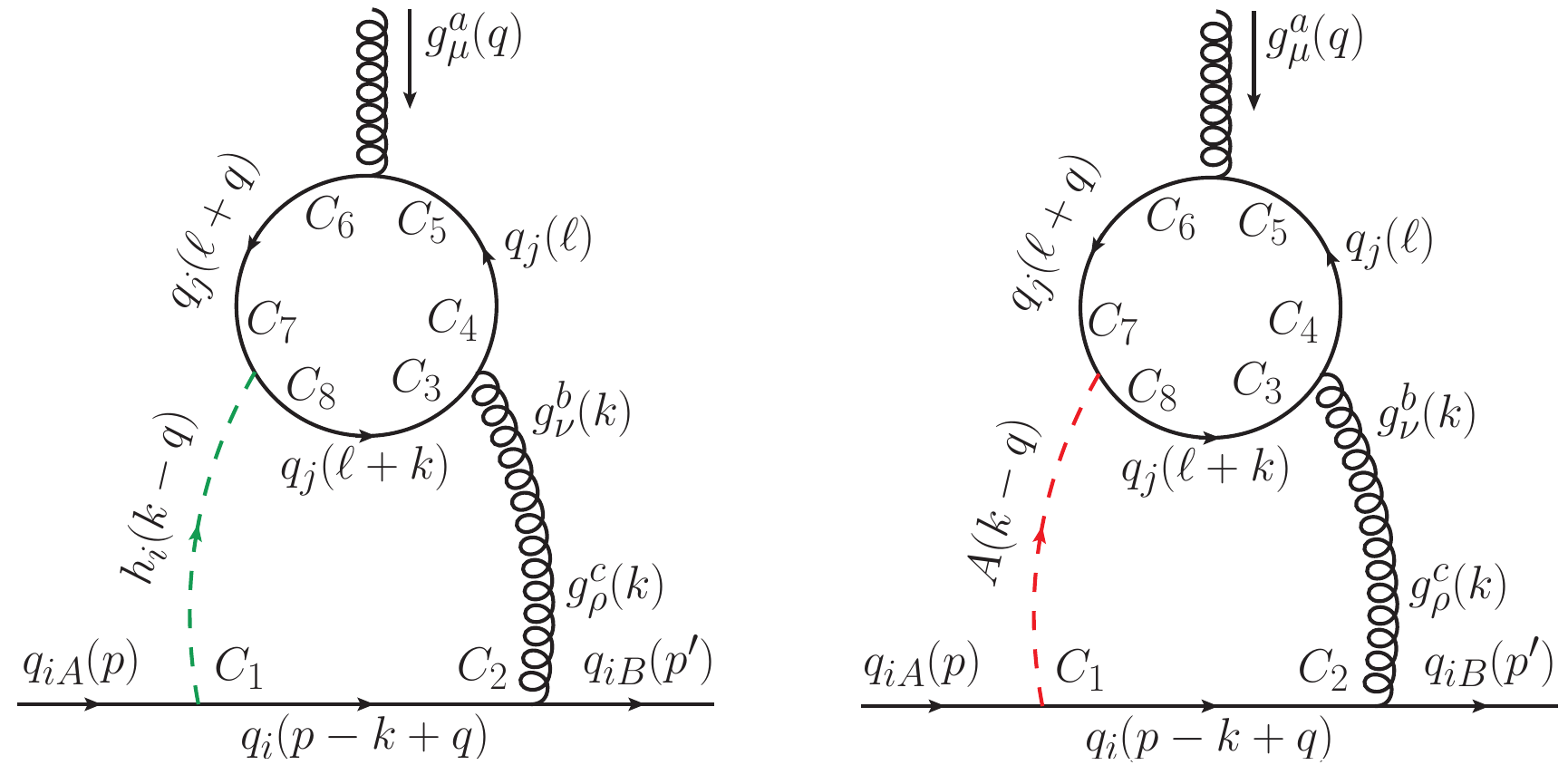}\\
	\,\,(a)\hspace{7.0cm}(b)
	\caption{Two-loop contributions to the CMDM of quarks are illustrated in Figures (a) and (b), depicting the contributions mediated by CP-even and CP-odd Higgs scalars, respectively.}
	\label{fig:cmdm2loop}
\end{figure}

This section details the two-loop Barr-Zee type contributions to the anomalous CMDM of quarks. Fig.~\ref{fig:cmdm2loop}a and \ref{fig:cmdm2loop}b display the corresponding two-loop Barr-Zee type diagrams with the momentum assignments, mediated by the CP-even and CP-odd scalars, respectively. The charged Higgs-mediated diagrams do not appear in this case. In calculating these two-loop diagrams, we follow the methods described in Ref.~\cite{Barr:1990vd, Bisal:2022nbn}. In this case, additionally, we treat the external gluon as off-shell during the calculation.

\begin{figure}[H]
	\centering
	\includegraphics[width=0.9\linewidth]{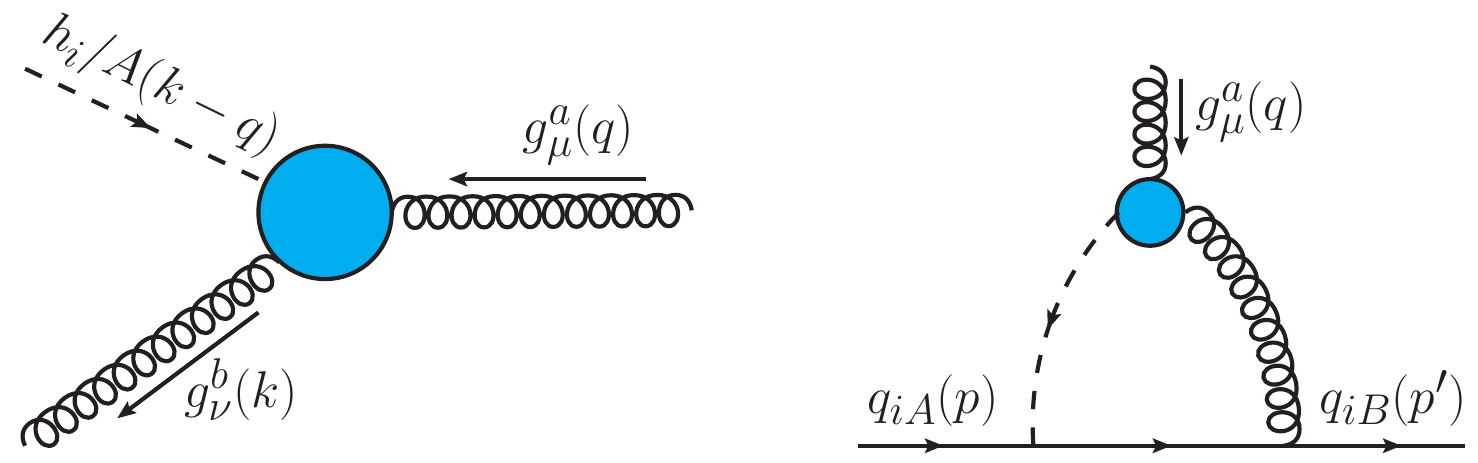}\\
	(a)\hspace{7.9cm}(b)
	\caption{(a) One-loop $h_i/Agg$ vertex which is used to calculate the second loop. (b) $qqg$ vertex where the blob represents the $h_i/Agg$ vertex.}
	\label{fig:effectivevertex}
\end{figure}

The full two-loop amplitude for Fig.~\ref{fig:cmdm2loop}a can be written as 

\begin{align}
\mathcal{M}_{(a),\,\rm 2L}^{\mu} =& \int\frac{d^dk}{(2\pi)^d}\int\frac{d^d\ell}{(2\pi)^d}\bar{u}(p^\prime)\Bigg[\big(-ig_sT^c_{C_2B}\gamma^\rho\big)\Biggl\{\frac{i\big(\slashed{p}-\slashed{k}+\slashed{q}+m_{q_i}\big)}{\big(p-k+q\big)^2-m_{q_i}^2}\delta_{C_1C_2}\Biggr\}\big(-iy_{q_i q_i h_i}\delta_{C_1A}\big)\Biggl\{\frac{i}{\big(k-q\big)^2-m_{h_i}^2}\Biggr\}\nonumber\\
&\times\Biggl\{-{\rm Tr}\Bigg[\big(-ig_sT^a_{C_5C_6}\gamma^\mu\big)\Biggl\{\frac{i\big(\slashed{\ell}+m_{q_j}\big)}{\ell^2-m_{q_j}^2}\delta_{C_4C_5}\Biggr\}\big(-ig_sT^b_{C_3C_4}\gamma^\nu\big)\Biggl\{\frac{i\big(\slashed{\ell}+\slashed{k}+m_{q_j}\big)}{\big(\ell+k\big)^2-m_{q_j}^2}\delta_{C_8C_3}\Biggr\}\big(-iy_{q_j q_j h_i}\delta_{C_7C_8}\big)\nonumber\\
&\times\Biggl\{\frac{i\big(\slashed{\ell}+\slashed{q}+m_{q_j}\big)}{\big(\ell+q\big)^2-m_{q_j}^2}\delta_{C_6C_7}\Biggr\}\Bigg]\Biggr\}\Bigg(\frac{-ig_{\nu\rho}}{k^2}\delta_{bc}\Bigg)\Bigg]u(p)~.
\label{Eq:2loopamp}
\end{align}

Now, we can recast Eq.~\eqref{Eq:2loopamp} in the following form:
\begin{align}
\mathcal{M}_{(a),\,\rm 2L}^{\mu}= Q_1\int\frac{d^dk}{(2\pi)^d}\bar{u}(p^\prime)\Bigg[\Biggl\{\frac{\gamma_\nu\big(\slashed{p}-\slashed{k}+\slashed{q}+m_{q_i}\big)}{D_1D_2D_3}\Biggr\}\big(\Sigma^{\mu\nu}\big)\Bigg]u(p)~,
\label{Eq:7redef}
\end{align}

where $Q_1=-iy_{q_i q_i h_i}g_sT^c_{C_2B}\delta_{C_1C_2}\delta_{C_1A}\delta_{bc}= -iy_{q_i q_i h_i}g_sT_{AB}^b$ and $D_1= k^2$, $D_2= \big(k-q\big)^2-m_{h_i}^2$, and $D_3=\big(p-k+q\big)^2-m_{q_i}^2$. Also, we have defined $\Sigma^{\mu\nu}$ in the following way:

\begin{align}
\Sigma^{\mu\nu} = Q_2\int \frac{d^d\ell}{(2\pi)^d}\Bigg[\Bigg(\frac{1}{D_4D_5D_6}\Bigg){\rm Tr}\big[\gamma^\mu\big(\slashed{\ell}+m_{q_j}\big)\gamma^\nu\big(\slashed{\ell}+\slashed{k}+m_{q_j}\big)\big(\slashed{\ell}+\slashed{q}+m_{q_j}\big)\big]\Bigg]~,
\label{Eq:1stloop}
\end{align}

where $Q_2= - y_{q_j q_j h_i}g_s^2 T^a_{C_5C_6}\delta_{C_4C_5}T^b_{C_3C_4}\delta_{C_8C_3}\delta_{C_7C_8}\delta_{C_6C_7}=-y_{q_j q_j h_i}g_s^2{\rm Tr}\big[T^aT^b\big]$ = $-\frac{1}{2}y_{q_j q_j h_i}g_s^2\delta_{ab}$ \big(using the relation ${\rm Tr}\big[T^aT^b\big]=\frac{1}{2}\delta^{ab}$\big) and $D_4=\ell^2-m_{q_j}^2$, $D_5=\big(\ell+k\big)^2-m_{q_j}^2$, and $D_6=\big(\ell+q\big)^2-m_{q_j}^2$. Here, we see that the loop integral $\Sigma^{\mu\nu}$ corresponds to the one-loop $h_igg$ vertex shown in Fig.~\ref{fig:effectivevertex}. First, we compute the integral $\Sigma^{\mu\nu}$ with respect to the loop momentum variable ``$\ell$" and then use it to compute the second loop integral in Eq.~\eqref{Eq:7redef}.
With the introduction of the Feynman parameters, Eq.~\eqref{Eq:1stloop} takes on the following form:

\begin{align}
\Sigma^{\mu\nu} = Q_2\Big[\mathcal{F}_1 g^{\mu\nu} + \mathcal{F}_2 k^\mu q^\nu + \mathcal{F}_3 k^\nu q^\mu + \mathcal{F}_4 k^{\mu} k^{\nu} + \mathcal{F}_5 q^\mu q^\nu\Big].
\label{Eq:9formfac}
\end{align}
The form factors $\mathcal{F}_1,...,\mathcal{F}_5$ are given by

\begin{align}
\mathcal{F}_1 =& \frac{i}{\big(4\pi\big)^2}\int_{0}^{1}dx\int_{0}^{1-x}dy\,\, 4m_{q_j}\Bigg[1 - m_{q_j}^2\Bigg(\frac{1}{\mathcal{D}}\Bigg) +\Bigg(\frac{1}{\mathcal{D}}\Bigg)\Bigl\{y^2k^2 + \big(2xy-1\big)\big(k.q\big)+ x^2 q^2\Bigr\}\Bigg]~,\\
\mathcal{F}_2 =& \frac{i}{\big(4\pi\big)^2}\int_{0}^{1}dx\int_{0}^{1-x}dy\,\,4m_{q_j} \Bigg[\Bigg(\frac{1}{\mathcal{D}}\Bigg)\big(1-4xy\big)\Bigg]~,\\
\mathcal{F}_3 =& \frac{i}{\big(4\pi\big)^2}\int_{0}^{1}dx\int_{0}^{1-x}dy\,\,4m_{q_j}\Bigg[\Bigg(-\frac{1}{\mathcal{D}}\Bigg)\big(1-2x\big)\big(1-2y\big)\Bigg]~,\\
\mathcal{F}_4 =& \frac{i}{\big(4\pi\big)^2}\int_{0}^{1}dx\int_{0}^{1-x}dy\,\,8m_{q_j}\Bigg[\Bigg(\frac{1}{\mathcal{D}}\Bigg)y\big(1-2y\big)\Bigg]~,\\
\mathcal{F}_5 =& \frac{i}{\big(4\pi\big)^2}\int_{0}^{1}dx\int_{0}^{1-x}dy\,\,8m_{q_j}\Bigg[\Bigg(\frac{1}{\mathcal{D}}\Bigg)x\big(1-2x\big)\Bigg]~,
\end{align}
where $\mathcal{D}= x(x-1)q^2 + y(y-1)k^2 + 2xy(k.q)+ m_{q_j}^2$, $x$ and $y$ are the Feynman parameters, respectively. Now we put the expressions for $\mathcal{F}_1,...,\mathcal{F}_5$ in Eq.~\eqref{Eq:7redef} and integrate over the loop momentum variable ``$k$". There are five different integrals corresponding to the five form factors in Eq.~\eqref{Eq:9formfac}. Finally, we pick up the coefficient of $\sigma^{\mu\nu}q_{\nu}$, which leads to the following integrals:

\begin{align}
\mathcal{I}_1 =&\,\,i\frac{m_{q_j}g_sQ}{\big(4\pi\big)^4} \int_{0}^{1}dx\int_{0}^{1-x}dy \Bigg[-\frac{2m_{q_j}^2}{y\big(y-1\big)^2}\Bigl\{\big(y-1\big)\mathbf{D}_0-x\,\mathbf{D}_3 + \big(y-1\big)\big(\mathbf{D}_1+\mathbf{D}_2\big)\Bigr\}+ 2\big(\widetilde{\mathbf{C}}_0+\widetilde{\mathbf{C}}_1+\widetilde{\mathbf{C}}_2\big)\nonumber\\
&+\frac{1}{y\big(y-1\big)}\Biggl\{m_{h_i}^2\big(1-2xy\big)\big(\mathbf{D}_0+\mathbf{D}_1+\mathbf{D}_2\big) + \big(1-2xy\big)\overline{\mathbf{C}}_0 + \big(1-2xy-2y^2\big)\mathbf{C}_2+ \big(1-2xy\big)\overline{\mathbf{C}}_1+q^2\nonumber\\
&\times\big(2xy+2x^2-1\big)\mathbf{D}_0 - \frac{x}{y-1}\Bigl\{\big(2y^2+2xy-1\big)\mathbf{C}_2+ \big(1-2xy\big)\overline{\mathbf{C}}_2+ \bigl\{\big(1-2xy\big)m_{h_i}^2+q^2\big(2x^2+2xy\nonumber\\
&-1\big)\bigr\}\mathbf{D}_3\Bigr\} + q^2\big(2x^2+2xy-1\big)\big(\mathbf{D}_1+\mathbf{D}_2\big)\Biggr\}\Bigg]~,
\label{eq:twoloop1}
\end{align}

\begin{align}
\mathcal{I}_2 = i\frac{m_{q_j}g_sQ}{\big(4\pi\big)^4} \int_{0}^{1}dx\int_{0}^{1-x}dy \Bigg[\frac{4xy-1}{y\big(y-1\big)^2}\Bigl\{-2x\xi\mathbf{D}_{23} + \big(y-1\big)\bigl\{q^2\big(\mathbf{D}_{12}+\mathbf{D}_{22}\big)-2\mathbf{D}_{00}\bigr\} + q^2\big(y-1\big)\mathbf{D}_2\Bigr\}\Bigg]~,
\label{eq:twoloop2}
\end{align}
where $Q=4iy_{q_i q_i h_i} y_{q_j q_j h_i} g_s^2T^{b}_{AB}\delta_{ab}=4iy_{q_i q_i h_i} y_{q_j q_j h_i}g_s^2T^{a}_{AB}$ and the PV functions are defined as (according to the convention of {\sc Package-X}) 
\begin{align}
\mathbf{D}_k =& \mathbf{D}_k\Bigg[q^2, m_{q_i}^2, m_{q_i}^2+q^2\Biggl\{\frac{x^2}{\big(y-1\big)^2}+1\Biggr\}+ 2\Biggl\{\frac{x}{y-1}-1\Biggr\}\xi, \frac{q^2x^2}{\big(y-1\big)^2}; m_{q_i}^2+q^2-2\xi, q^2\Biggl\{\frac{x^2}{\big(y-1\big)^2}+1\Biggr\}\nonumber\\
&+\frac{4x\xi}{y-1};0,m_{h_i},m_{q_i},M\Bigg]~,\nonumber\\
\mathbf{D}_{k\ell} =& \mathbf{D}_{k\ell}\Bigg[q^2, m_{q_i}^2, m_{q_i}^2+q^2\Biggl\{\frac{x^2}{\big(y-1\big)^2}+1\Biggr\}+ 2\Biggl\{\frac{x}{y-1}-1\Biggr\}\xi, \frac{q^2x^2}{\big(y-1\big)^2}; m_{q_i}^2+q^2-2\xi, q^2\Biggl\{\frac{x^2}{\big(y-1\big)^2}+1\Biggr\}\nonumber\\
&+\frac{4x\xi}{y-1};0,m_{h_i},m_{q_i},M\Bigg]~,\nonumber
\end{align}
\begin{align}
\mathbf{C}_k =& \mathbf{C}_k\Bigg[m_{q_i}^2, m_{q_i}^2+q^2\Biggl\{\frac{x^2}{\big(y-1\big)^2}+1\Biggr\}+ 2\Biggl\{\frac{x}{y-1}-1\Biggr\}\xi, q^2\Biggl\{\frac{x^2}{\big(y-1\big)^2}+1\Biggr\}+\frac{4x\xi}{y-1}; m_{h_i}, m_{q_i}, M\Bigg]~,\nonumber\\
\overline{\mathbf{C}}_k =& \mathbf{C}_k\Bigg[m_{q_i}^2+q^2-2\xi, m_{q_i}^2+q^2\Biggl\{\frac{x^2}{\big(y-1\big)^2}+1\Biggr\}+ 2\Biggl\{\frac{x}{y-1}-1\Biggr\}\xi, \frac{q^2x^2}{\big(y-1\big)^2};0,m_{q_i},M\Bigg]~,\nonumber\\
\widetilde{\mathbf{C}}_k =& \mathbf{C}_k\big[m_{q_i}^2+q^2-2\xi, m_{q_i}^2, q^2; 0, m_{q_i}, m_{h_i}\big]~. \nonumber
\end{align}
Also, the following definitions have been used:
\begin{align}
\xi= m_{q_i}^2+\frac{1}{2}\big(q^2-2m_{q_i}^2\big)~,\hspace{1cm} M=\Bigg[\frac{m_{q_j}^2}{y\big(1-y\big)}+ \frac{q^2x\big(x+y-1\big)}{y\big(1-y\big)^2}\Bigg]^{\frac{1}{2}}~.
\label{Eq:defXiM}
\end{align}

There are no such integrals corresponding to the form factors $\mathcal{F}_3$ and $\mathcal{F}_5$, i.e., the coefficient of $\sigma^{\mu\nu}q_\nu$ is zero for these cases. Also, the integral corresponding to the form factor $\mathcal{F}_4$ is evaluated to zero. 

Now, we focus on the diagram in Fig.~\ref{fig:cmdm2loop}b. In this case, we follow the same procedure as in the previous one.
The only difference is that the Yukawa couplings contain a $\gamma^5$-term. In this case, we can write similar to Eq.~\eqref{Eq:7redef} as

\begin{align}
\mathcal{M}_{(b),\,\rm 2L}^{\mu}= Q_1^\prime\int\frac{d^dk}{(2\pi)^d}\bar{u}(p^\prime)\Bigg[\Biggl\{\frac{\gamma_\nu\big(\slashed{p}-\slashed{k}+\slashed{q}+m_{q_i}\big)\gamma^5}{D_1D_2^{\prime}D_3}\Biggr\}\big(\widetilde{\Sigma}^{\mu\nu}\big)\Bigg]u(p)~,
\label{Eq:18redefA}
\end{align} 
where $Q_1^\prime = -iy_{q_i q_i A}g_sT^b_{AB}$, $D_2^{\prime} = \big(k-q\big)^2-m_A^2$, and $D_1$ and $D_3$ are the same as defined earlier. In this case, the $Agg$ one-loop vertex can be written as 

\begin{align}
\widetilde{\Sigma}_{\mu\nu} = Q_2^\prime \widetilde{\mathcal{F}} \epsilon_{\mu\nu\rho\sigma} k^{\rho}q^{\sigma}~,
\label{Eq:pseudoscalar}
\end{align}
where $Q_2^\prime = -\frac{1}{2}y_{q_j q_j A}g_s^2 \delta_{ab}$ and the form factor $\widetilde{\mathcal{F}}$ is given by
\begin{align}
\widetilde{\mathcal{F}} = \frac{1}{\big(4\pi\big)^2}\int_{0}^{1}dx\int_{0}^{1-x}dy\Bigg[\frac{4m_{q_j}}{\mathcal{D}}\Bigg]~,
\label{Eq:Ftilde}
\end{align}
where the denominator $\mathcal{D}$ is defined earlier. Now, using  Eq.~\eqref{Eq:pseudoscalar} and \eqref{Eq:Ftilde}, we can calculate the integral in Eq.~\eqref{Eq:18redefA}. Finally, collecting the coefficient of $\sigma^{\mu\nu}q_\nu$, we obtain

\begin{align}
\widetilde{\mathcal{I}} = i\frac{m_{q_j}g_sQ^\prime}{\big(4\pi\big)^4}\int_{0}^{1}dx \int_{0}^{1-x}dy \, \big(4\mathbf{D}_{00}\big)~,
\end{align}
where $Q^\prime = 4iy_{q_i q_i A} y_{q_j q_j A} g_s^2T^a_{AB}$ and the PV function $\mathbf{D}_{00}$ is defined as
\begin{align*}
\mathbf{D}_{00} =& \mathbf{D}_{00} \Bigg[q^2, m_{q_i}^2, m_{q_i}^2+q^2\Biggl\{\frac{x^2}{\big(y-1\big)^2}+1\Biggr\}+ 2\Biggl\{\frac{x}{y-1}-1\Biggr\}\xi, \frac{q^2x^2}{\big(y-1\big)^2}; m_{q_i}^2+q^2-2\xi, q^2\Biggl\{\frac{x^2}{\big(y-1\big)^2}+1\Biggr\}\\
&+\frac{4x\xi}{y-1};0,m_{A},m_{q_i},M\Bigg]~,
\end{align*}
where the definitions of $\xi$ and $M$ are given in Eq.~\eqref{Eq:defXiM}.

Thus, we can write the two-loop corrections to the CMDM from the diagrams in Fig.~\ref{fig:cmdm2loop}a and \ref{fig:cmdm2loop}b as

\begin{align}
    \hat{\mu}_{q_i}^{(a),\, \rm 2L} =&\, \frac{m_{q_i}}{g_s}\big(\mathcal{I}_1 + \mathcal{I}_2\big)~,\\
    \hat{\mu}_{q_i}^{(b),\, \rm 2L} =&\, \frac{m_{q_i}}{g_s}\widetilde{\mathcal{I}}~.
\end{align}

\section{The Aligned Two-Higgs-Doublet Model}
\label{sec:modelTHDM}
Let's consider extending the SM by introducing a second complex scalar doublet with hypercharge $Y=\frac{1}{2}$. In general, the neutral components of both doublets ($\Phi_1$, $\Phi_2$) acquire vacuum expectation values (vev). However, it is convenient to work on the so-called Higgs basis, where only one doublet acquires vev.

\begin{align}
\Phi_1=    \begin{bmatrix}
    G^+ \\
    \frac{1}{\sqrt{2}} \big(v+S_1+iG^0\big) \\
\end{bmatrix}~,
\hspace{2cm}\Phi_2=    \begin{bmatrix}
    H^+ \\
    \frac{1}{\sqrt{2}} \big(S_2+iS_3\big) \\
\end{bmatrix}~,
\end{align}
where $G^\pm$ and $G^0$ are the charged and neutral Goldstone fields. Here $\Phi_1$ acts as an SM scalar doublet with vev $(v=\sqrt{2}G_F)^{-1/2}\approx246$ GeV. The scalar sector encompasses five degrees of freedom: two charged fields $H^\pm(x)$ and three neutral fields $\phi_i^0(x)=\bigl\{h(x), H(x), A(x) \bigr\}$, which can be obtained by the orthogonal transformation of the $S_i$ fields $\phi_i^0(x)=\mathcal{R}_{ij} S_j(x)$. 

The most general scalar potential, which is invariant under $SU(2)_L \otimes U(1)_Y$ can be expressed as
\begin{align}
    V=&\mu_1 \Phi_1^\dagger\Phi_1 + \mu_2\Phi_2^\dagger\Phi_2 + \Big[\mu_3 \Phi_1^\dagger\Phi_2 + \mu_3^* \Phi_2^\dagger \Phi_1\Big] + \frac{1}{2}\lambda_1 \big(\Phi_1^\dagger\Phi_1\big)^2 + \frac{1}{2}\lambda_2 \big(\Phi_2^\dagger\Phi_2\big)^2 + \lambda_3 \big(\Phi_1^\dagger\Phi_1\big)\big(\Phi_2^\dagger\Phi_2\big)\nonumber\\
    &+ \lambda_4 \big(\Phi_1^\dagger\Phi_2\big)\big(\Phi_2^\dagger\Phi_1\big) + \Big[\Big(\frac{1}{2}\lambda_5\Phi_1^\dagger\Phi_2+ \lambda_6\Phi_1^\dagger\Phi_1+\lambda_7\Phi_2^\dagger\Phi_2\Big)\big(\Phi_1^\dagger\Phi_2\big) + {\rm h.c.}\Big]~,
\end{align}
where all the parameters are real except $\mu_3$, $\lambda_5$, $\lambda_6$, and $\lambda_7$. The minimization of the potential yields the following relations:
\begin{align}
    \mu_1=-\frac{1}{2}\lambda_1 v^2~, \hspace{1cm} \mu_3=-\frac{1}{2}\lambda_6 v^2~. 
\end{align}
Furthermore, one can reabsorb one phase into $\Phi_2$. Additionally, for simplifying the analysis, we assume the CP-conserving potential, consequently reducing the total degrees of freedom to nine.
The physical scalar masses are determined by the quadratic terms in the potential~\cite{Celis:2013rcs}.
\begin{align}
    m_{H^\pm}^2=&\mu_2+ \frac{1}{2}\lambda_3 v^2~, \hspace{1.5cm}m_A^2 = m_{H^\pm}^2 +\frac{1}{2} \big(\lambda_4-\lambda_5\big)v^2~,\nonumber\\
    m_h^2 =&\frac{1}{2}\big(\Sigma-\Delta\big)~, \hspace{1.73cm} m_H^2 = \frac{1}{2}\big(\Sigma+\Delta\big)~,
\end{align}
where $\Sigma = m_{H^\pm}^2 + \frac{1}{2}v^2 (2\lambda_1+\lambda_4+\lambda_5)$ and $\Delta=\Big[\Bigl\{m_{H^\pm}^2 + \frac{1}{2}v^2 (-2\lambda_1+\lambda_4+\lambda_5)\Bigr\}^2 + 4v^4 \lambda_6^2\Big]^{1/2}$. For the CP symmetric scalar potential, the admixture between CP-odd and CP-even fields vanishes. Then, $A(x)=S_3(x)$ and the CP-even neutral mass eigenstate are the linear combinations of $S_1(x)$ and $S_2(x)$:
\begin{align}
    \begin{pmatrix}
     h \\
     H \\
\end{pmatrix}
=\begin{pmatrix}
    \cos\tilde{\alpha} & \sin\tilde{\alpha} \\
    -\sin\tilde{\alpha} & \cos\tilde{\alpha} \\
\end{pmatrix}
\begin{pmatrix}
     S_1 \\
     S_2 \\
\end{pmatrix}~.
\end{align}
The sign of $\sin\tilde{\alpha}$ can be fixed by performing a phase redefinition of the neutral CP-even scalar fields.

The most generic Yukawa Lagrangian with the SM fermionic content leads to Flavor-Changing Neutral Currents (FCNCs) due to the inability to simultaneously diagonalize the fermionic couplings of the two scalar doublets in flavor space. One can eliminate the non-diagonal neutral couplings by necessitating the alignment in flavor space of the Yukawa matrices~\cite{Pich:2009sp}. The three proportionality parameters, $\zeta_f$ ($f=u,d,\ell$), are arbitrary complex numbers, introducing the new sources of CP violation within them.

The Yukawa interactions of the ATHDM, expressed in terms of the fermion mass eigenstate, can be written as~\cite{Pich:2009sp, Manohar:2006ga, Pich:2010ic}
\begin{align}
    \mathcal{L}_Y \supset -\frac{\sqrt{2}}{v} H^+ \Bigl\{\bar{u}\big[\zeta_dVM_d\mathbf{P}_R - \zeta_u M_u^\dagger V\mathbf{P}_L\big]d + \zeta_\ell \bar{\nu}M_\ell \mathbf{P}_R \ell\Bigr\}-\frac{1}{v}\sum_{\phi_i^0, f} y_f^{\phi_i^0}\phi_i^0\big[\bar{f}M_f\mathbf{P}_R f\big]+{\rm h.c.}~,
\end{align}
where $\mathbf{P}_{R, L}=\frac{1}{2}\big(1\pm \gamma^5\big)$ are the right-handed and left-handed chirality operators, $M_f$ is the diagonal fermion mass matrices, and the couplings of the neutral scalar fields can be cast as:
\begin{align}
   y_f^h=& \cos\tilde{\alpha} + \zeta_f\sin\tilde{\alpha}~,\hspace{1.5cm} y_{d,\ell}^A=i\zeta_{d,\ell}~,~\nonumber\\
   y_f^H=& -\sin\tilde{\alpha} + \zeta_f\cos\tilde{\alpha}~,\hspace{1.35cm} y_u^A= -i\zeta_u~,
\end{align}
where we have considered the CP-conserving scalar potential and the real alignment parameter $\zeta_f$.

The conventional THDM scenarios based on $\mathcal{Z}_2$ symmetries are the special cases of the ATHDM framework. Those models can be retrieved by imposing the conditions $\mu_3=\lambda_6=\lambda_7=0$ along with the following conditions:
\begin{align}
    &{\rm Type-I:}\hspace{0.2cm} \zeta_u=\zeta_d=\zeta_\ell= \cot\beta~, \hspace{1.5cm}{\rm Type-II:}\hspace{0.2cm} \zeta_u=-\frac{1}{\zeta_d}=-\frac{1}{\zeta_\ell}= \cot\beta~,\nonumber\\
    &{\rm Type-X:}\hspace{0.2cm} \zeta_u=\zeta_d=-\frac{1}{\zeta_\ell}= \cot\beta~, \hspace{1.0cm}{\rm Type-Y:}\hspace{0.2cm} \zeta_u=-\frac{1}{\zeta_d}=\zeta_\ell= \cot\beta~,
    \label{eq:convTHDMs}
\end{align}

\section{Results and Discussions}
\label{sec:results}
This section presents the numerical results for the CMDM of the top quark at both one-loop and two-loop levels. First, we present the results for the ATHDM and then for the conventional THDMs with $\mathcal{Z}_2$ symmetries. All these conventional THDMs can be obtained as special cases of the more general ATHDM for different values of the alignment parameters $\zeta_{u,d,\ell} $ as given in Eq.~\eqref{eq:convTHDMs}. In this analysis, we take the ``light scenario" as given in Ref.~\cite{Eberhardt:2020dat} where the lightest CP-even scalar ($h$) is the SM-like Higgs with $m_h\simeq 125~$GeV, and other scalars ($H, H^\pm$, A) lie above it.

\subsection{The CMDM of the top quark in the ATHDM}
\label{sec:ATHDMcmdm}

 Here, we take the variation of the alignment parameter in the range $\zeta_t\in [-1.5, 1.5]$.  The authors of Ref.~\cite{Eberhardt:2020dat} compare the alignment parameter constraints from flavor observables with 68\% and 95.5\% probability regions in the global fit. They have shown that the limits on $\zeta_d$ and $\zeta_\ell$ become stronger if one considers all the constraints. Also, 
 larger values of $\zeta_u$ require smaller values $\zeta_d$ and vice versa so that the product $\zeta_u\zeta_d$ becomes smaller. Since except for the one-loop diagram mediated by the charged Higgs (which is found to be much smaller compared to the other contributions, hence will not be included in the final results), all the one-loop and two-loop contributions do not depend on $\zeta_d$, we can take any value of $\zeta_u$ in the range [-1.5, 1.5], as evident from the global fit in Ref.~\cite{Eberhardt:2020dat}. This is also roughly consistent with the perturbativity requirement, $\big\lvert\big(\sqrt{2}\zeta_{u,d}m_{u,d}\big)/v\big\rvert \lesssim 1$ (for details, see Ref.~\cite{Allwicher:2021rtd}).

\begin{figure}[H]
	\centering
	\includegraphics[width=0.415\linewidth]{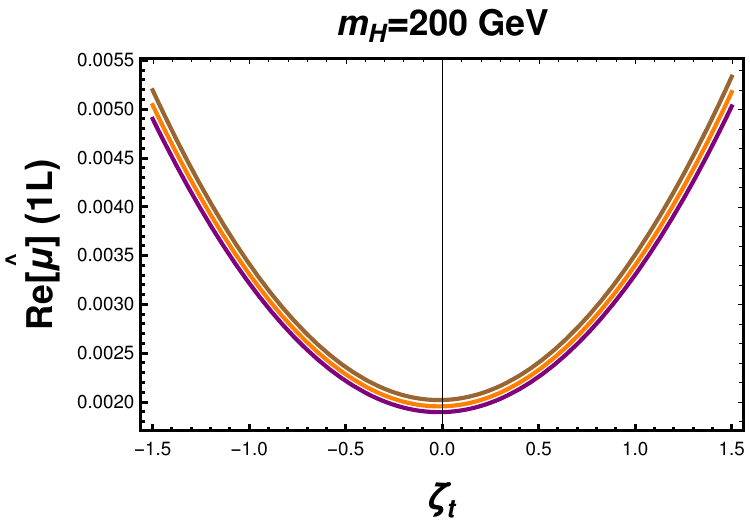}
    \includegraphics[width=0.525\linewidth]{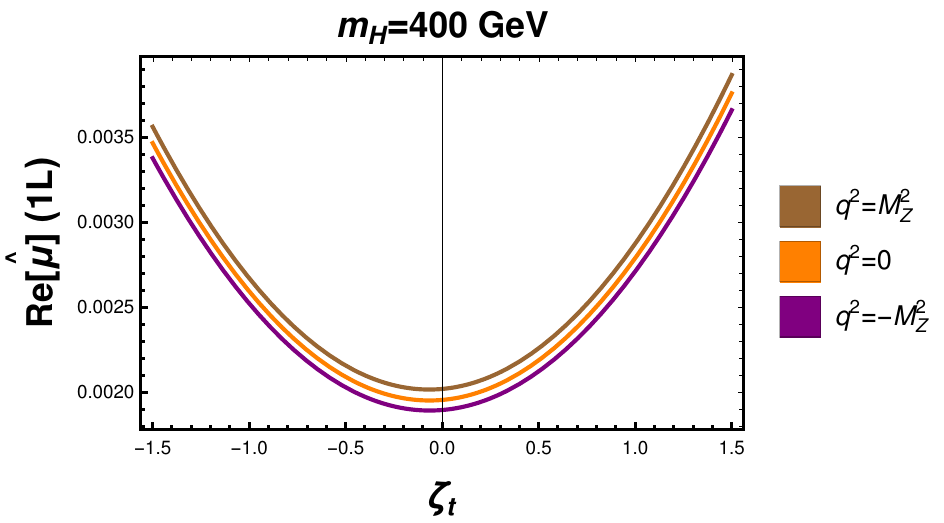}\\
    \hspace{-0.5cm}(a)\hspace{6.35cm}(b)\\
    \includegraphics[width=0.415\linewidth]{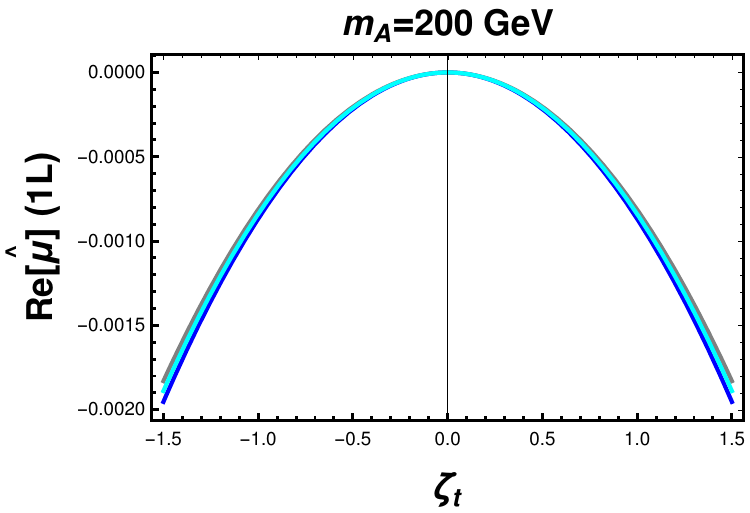}
    \includegraphics[width=0.525\linewidth]{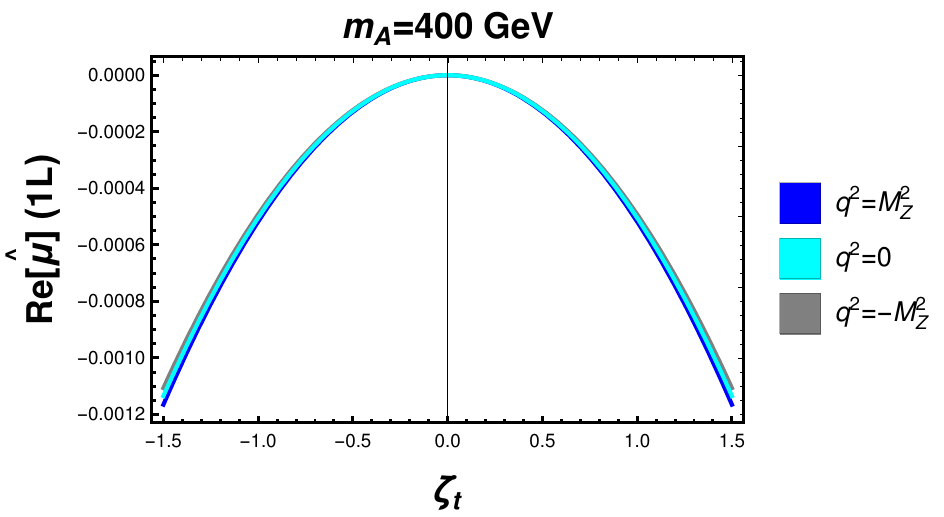}\\
    \hspace{-0.4cm}(c)\hspace{6.35cm}(d)
	\caption{One-loop contributions to the CMDM of the top quark in the ATHDM. Here, figures (a) and (b) show the variations of the one-loop contributions to the real part of the top quark CMDM mediated by the neutral CP-even Higgs scalars with the alignment parameter $\zeta_t$ for $m_H=200$~GeV and 400~GeV, respectively. Figures (c) and (d) show the same, except that the loop is mediated by the CP-odd scalar.}
	\label{plot:oneloopcmdm}
\end{figure}

Fig.~\ref{plot:oneloopcmdm} shows the one-loop contributions to the CMDM of the top quark in the ATHDM. Here, Fig.~\ref{plot:oneloopcmdm}a and \ref{plot:oneloopcmdm}b show the variations of the one-loop contributions to the real part of CMDM mediated by the neutral CP-even Higgs scalars $h, H$ (Fig.~\ref{fig:oneloopcmdm}a) with the alignment parameter $\zeta_t$ for $m_H=200$~GeV and 400~GeV, respectively. Note that the absorptive part of the CMDM is very small, and therefore, we have not shown it in the following plots. Here, the brown curve illustrates the value of Re[$\hat{\mu}$] at $q^2=M_Z^2$,  reaching its maximum value of $5.4\times 10^{-3}$ (Fig.~\ref{plot:oneloopcmdm}a) for $m_H=200$~GeV with the corresponding $\zeta_t=1.5$. Similarly, the orange and purple curves represent the values of CMDM at $q^2=0$ and $-M_Z^2$, respectively. For $m_H=400$~GeV, the maximum value of CMDM reduces to $3.9\times 10^{-3}$ (Fig.~\ref{plot:oneloopcmdm}b).

\begin{figure}[H]
	\centering
	\includegraphics[width=0.41\linewidth]{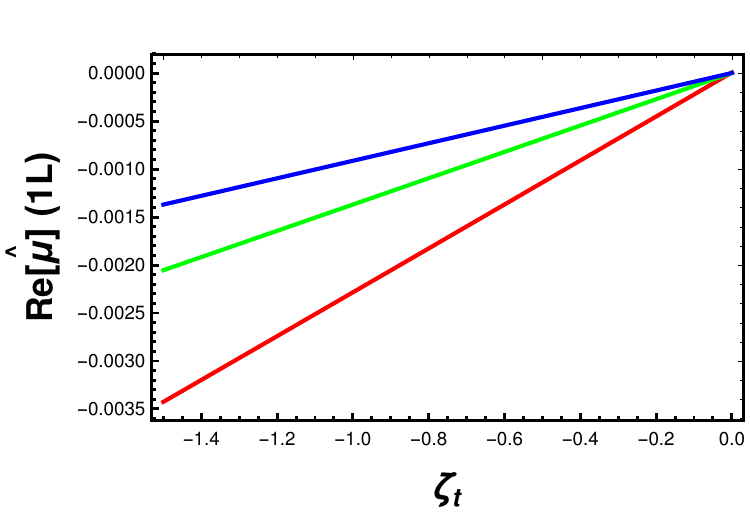}
    \includegraphics[width=0.56\linewidth]{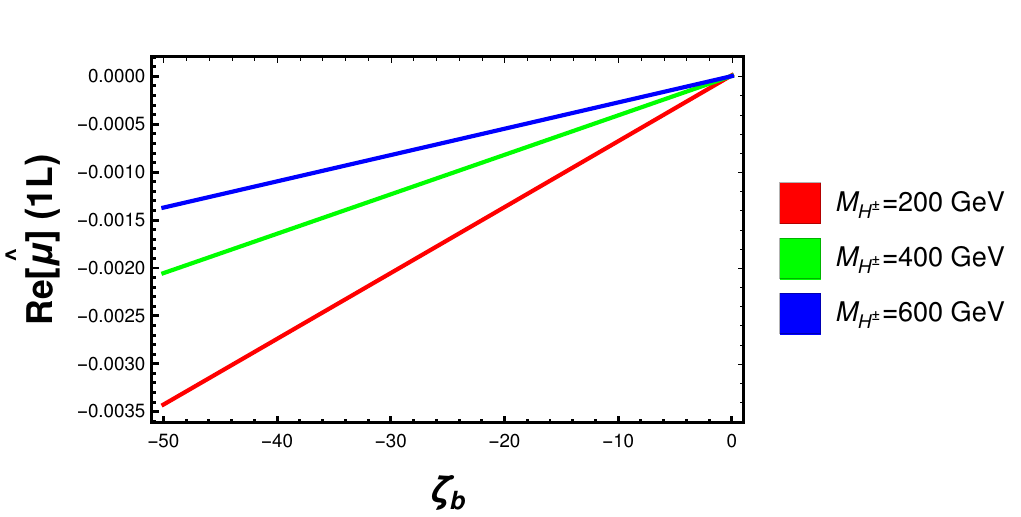}\\
    \hspace{-1.05cm}(a)\hspace{6.27cm}(b)\\
	\caption{One-loop contributions to the CMDM mediated by the charged Higgs. Here, (a) shows the variations of CMDM with $\zeta_t$ for a fixed value of $\zeta_b=-50$. Similarly, (b) shows the variations of the same with $\zeta_b$ for a fixed value of $\zeta_t=-1.5$. In both cases, the red, green, and blue curves correspond to the charged Higgs masses $m_{H^\pm}=200$~GeV, 400~GeV, and 600~GeV. respectively. Also, here, only the values of Re[$\hat{\mu}$] at $q^2=M_Z^2$ are shown.}
	\label{plot:oneloopHpm}
\end{figure}

Fig.~\ref{plot:oneloopcmdm}c and \ref{plot:oneloopcmdm}d depict the variations of the top quark CMDM with $\zeta_t$ for the pseudoscalar mediated one-loop diagram (Fig.~\ref{fig:oneloopcmdm}b). Here, the CMDM takes the negative sign and destructively interferes with the contributions mediated by the CP-even Higgs scalars. The blue, cyan, and gray contours correspond to the evaluation of the CMDM at $q^2=M_Z^2$, 0, and $-M_Z^2$, respectively.  The maximum value (in the negative sense) reaches up to $-2.0\times 10^{-3}$ for $m_A=200$~GeV.
In this case, the value of Re[$\hat{\mu}$] does not change much for the different values of the external gluon momentum transferred $q^2$. When the pseudoscalar mass increases to $m_A=400$~GeV, the CMDM decreases (in the negative sense) significantly to $\sim -1.2\times 10^{-3}$.
 The variations of CMDM with $\zeta_t$ for the charged Higgs-mediated one-loop diagram are shown in Fig.~\ref{plot:oneloopHpm}a and \ref{plot:oneloopHpm}b for $m_{H^\pm}=200$ GeV, 400 GeV, and 600 GeV, represented by the red, green, and blue curves, respectively. Here, only the values of the CMDM at $q^2=M_Z^2$ are shown. This diagram involves both the up-type and down-type alignment parameters corresponding to the top and bottom quarks. In Fig.~\ref{plot:oneloopHpm}a, the down-type alignment parameter, $\zeta_b$, is fixed at $-50$, and $\zeta_t$ has been varied in the range $\zeta_t\in [-1.5,0]$, which gives rise to $\zeta_u\zeta_d>0$.
 This is taken because of the fact that the parameter space for $\zeta_u\zeta_d<0$ is almost ruled out by the current data on the branching ratio of the weak radiative decay $\bar{B}\to X_s\gamma$~\cite{Jung:2010ik}. Here, the CMDM can take a maximum value of (in the negative sense) $-3.5\times 10^{-3}$ for $m_{H^\pm}=200$~GeV, which corresponds to $(\zeta_u, \zeta_d)=(-1.5, -50)$. This region is already excluded by different flavor observables~\cite{Eberhardt:2020dat}. The allowed parameters space corresponds to $|\zeta_u|\to 0$ (when $|\zeta_d|\to 50$), which results the value of CMDM being $\mathcal{O}\big(10^{-5}\big)$. Clearly, this is two orders of magnitude smaller than the neutral Higgs contribution. Fig.~\ref{plot:oneloopHpm}b shows the variation of CMDM with $\zeta_b$ for the fixed value of $\zeta_t=-1.5$.

\begin{figure}[H]
	\centering
	\includegraphics[width=0.425\linewidth]{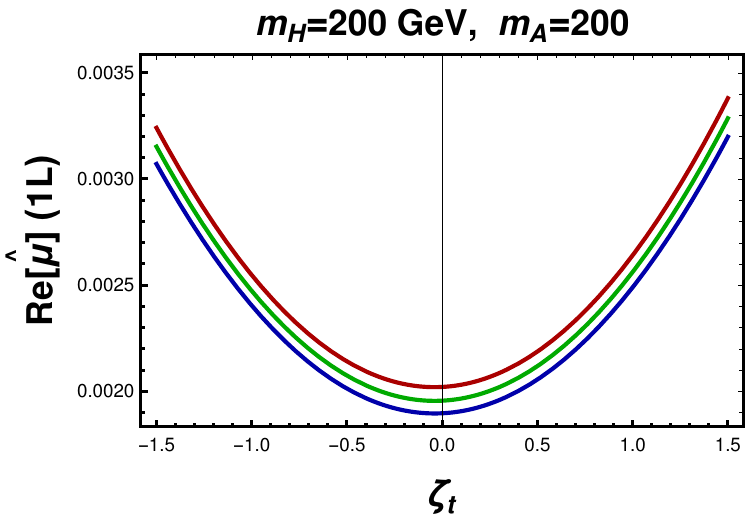}
    \includegraphics[width=0.530\linewidth]{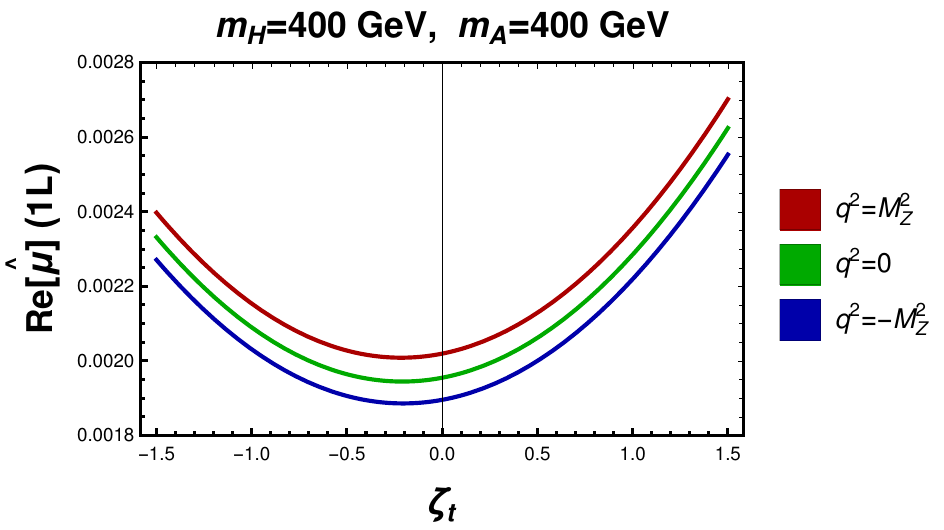}\\
    \hspace{-0.5cm}(a)\hspace{6.435cm}(b)\\
	\caption{Total one-loop corrections to the CMDM of the top quark in the ATHDM. Figures (a) and (b) show the variations of the one-loop CMDM with $\zeta_t$ for $m_{H, A}=200$~GeV and 400~GeV, respectively. Here, dark red, dark green, and dark blue colors correspond to the CMDM at $q^2=M_Z^2$, 0, and $-M_Z^2$, respectively.}
	\label{plot:oneloopATHDM}
\end{figure}

In Fig.~\ref{plot:oneloopATHDM}a and \ref{plot:oneloopATHDM}b, the total one-loop corrections to the top quark CMDM in the ATHDM are presented for $m_{H, A}=200$~GeV and 400~GeV, respectively. Here, dark red, dark green, and dark blue colors correspond to the CMDM evaluated at $q^2=M_Z^2$, 0, and $-M_Z^2$, respectively. The maximum value of the total one-loop CMDM at $q^2=M_Z^2$ is $3.4\times 10^{-3}$, which corresponds to $m_{H, A}=200$~GeV and $\zeta_t=1.5$.
 One important thing is that the CMDM receives its maximum value in the positive $\zeta_t$ region (the same thing happens in Fig.~\ref{plot:oneloopcmdm}a and \ref{plot:oneloopcmdm}b). The reason is that the Yukawa coupling of the top quark with light Higgs in the ATHDM is proportional to $(\cos\tilde{\alpha}+\zeta_t \sin\tilde{\alpha})$, and for the heavy Higgs, it is $(-\sin\tilde{\alpha}+\zeta_t \cos\tilde{\alpha})$. Here, the parameter $\tilde{\alpha}$ has a constraint from the global fit: $-0.04 \leq \tilde{\alpha} \leq 0.04$ (with a $95.5\%$ probability)~\cite{Eberhardt:2020dat}. 
 Here, we take the positive value of $\tilde{\alpha}$ to be 0.04\footnote{The equivalent solution arises from flipping the sign of $\tilde{\alpha}$, which corresponds to a sign flip in the coupling $\zeta_t$.}. When $\zeta_t$ takes the negative sign, the light Higgs Yukawa coupling decreases, leading to an overall decrease in the CMDM.
 But for positive $\zeta_t$, the light Higgs Yukawa coupling increases, which helps to increase the CMDM. Now, when $m_H=400~$GeV, the heavy Higgs contribution becomes smaller, and the light Higgs-mediated diagram dominates. Therefore, the Yukawa coupling effect of the light Higgs becomes more prominent, which raises the CMDM more in the positive $\zeta_t$ region.

\begin{figure}[H]
	\centering
	\includegraphics[width=0.43\linewidth]{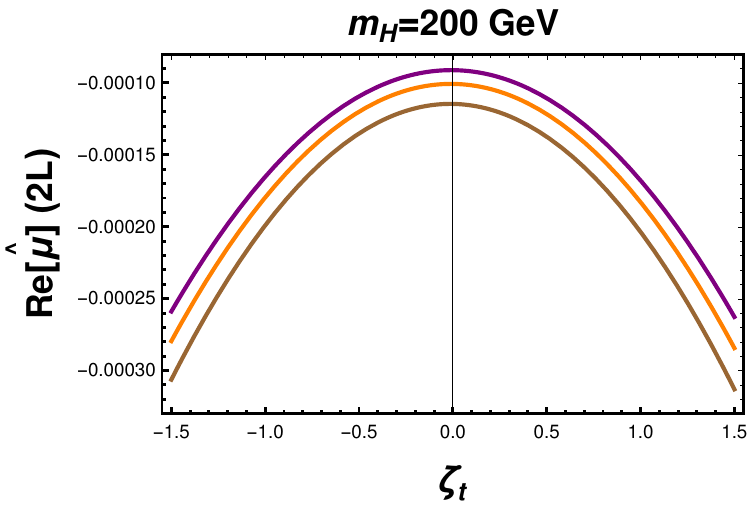}
    \includegraphics[width=0.54\linewidth]{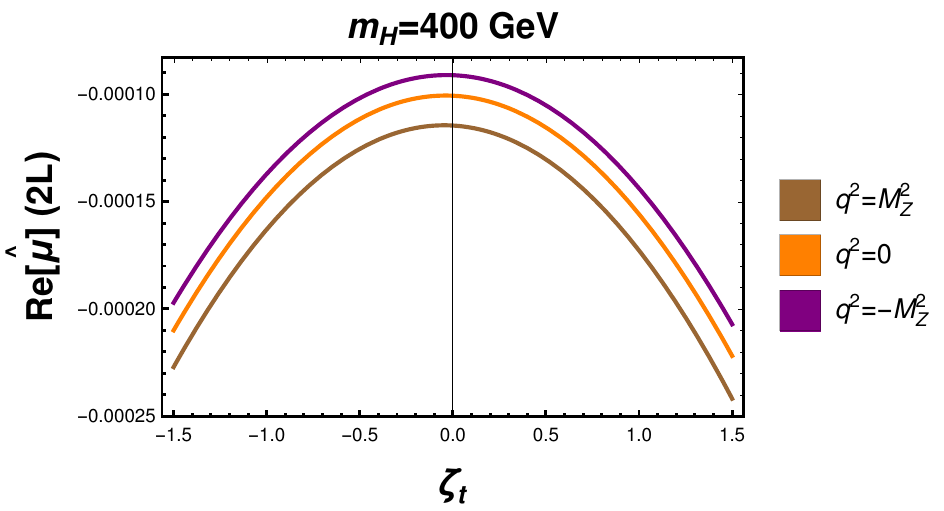}\\
    \hspace{-0.3cm}(a)\hspace{6.55cm}(b)\\
    \includegraphics[width=0.43\linewidth]{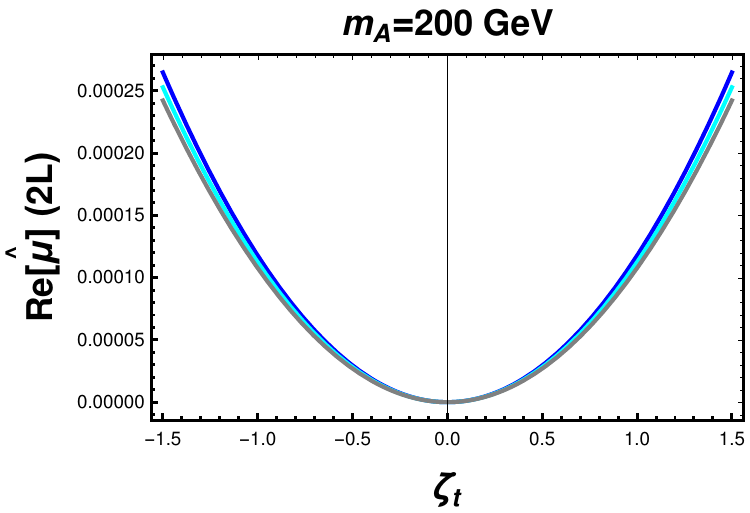}
    \includegraphics[width=0.54\linewidth]{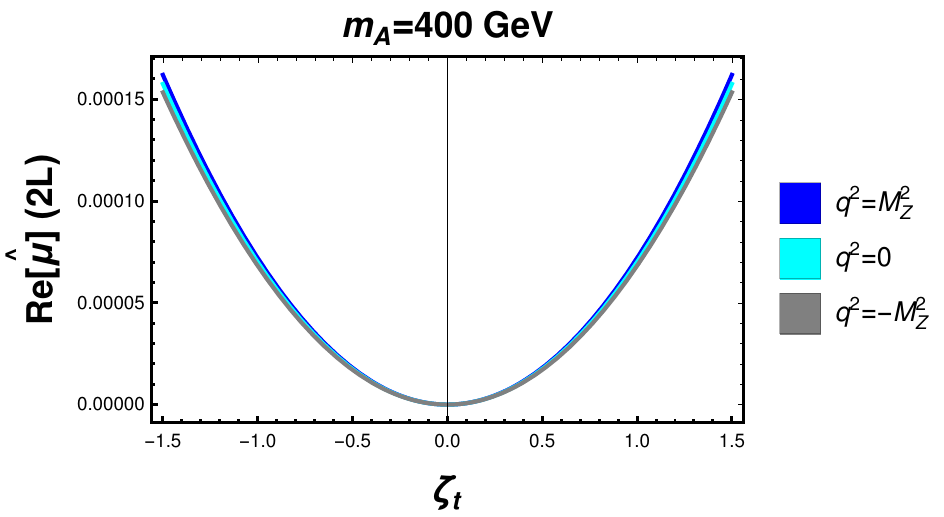}\\
    \hspace{-0.3cm}(c)\hspace{6.55cm}(d)
	\caption{Two-loop contributions to the CMDM of the top quark in the ATHDM. Here, (a) and (b) depict the variations of the CMDM with $\zeta_t$, mediated by the CP-even Higgs scalars, where the heavy Higgs masses are taken as $m_H=200$~GeV and 400~GeV, respectively. Figures (c) and (d) show the same for the CP-odd scalar-mediated diagram where $m_A=200$~GeV and 400~GeV are assumed.}
	\label{plot:twoloopcmdm}
\end{figure}

 Fig.~\ref{plot:twoloopcmdm} portrays the two-loop contributions to the CMDM of the top quark in the ATHDM. Fig.~\ref{plot:twoloopcmdm}a and \ref{plot:twoloopcmdm}b illustrate the variations of the CMDM with $\zeta_t$, mediated by the CP-even Higgs scalars, where the heavy Higgs masses are taken as $m_H = 200$ GeV and 400 GeV, respectively. The color codes are the same as the one-loop contributions plots. Here, unlike the one-loop contributions, the CP-even neutral Higgs contributions come with a negative sign, and the maximum value (in the negative sense) reaches up to $-3.1\times 10^{-4}$ at $q^2=M_Z^2$ for $m_H=200$~GeV. Clearly, this is one order of magnitude smaller than the corresponding one-loop contributions. For $m_H=400$~GeV, the value of CMDM reduces (in the negative sense) to $-2.3\times 10^{-4}$. Similarly, Fig.~\ref{plot:twoloopcmdm}c and \ref{plot:twoloopcmdm}d show the same for the CP-odd Higgs-mediated diagram, where $m_A=200$ GeV and 400 GeV are assumed.
 In this case, too, the two-loop contributions come with opposite signs compared to the one-loop contributions. Here, the maximum contributions can reach up to $2.6\times 10^{-4}$, which interfere destructively with the CP-even Higgs-mediated contributions at two-loop order. Therefore, the overall two-loop contributions decrease.

\begin{figure}[H]
	\centering
	\includegraphics[width=0.43\linewidth]{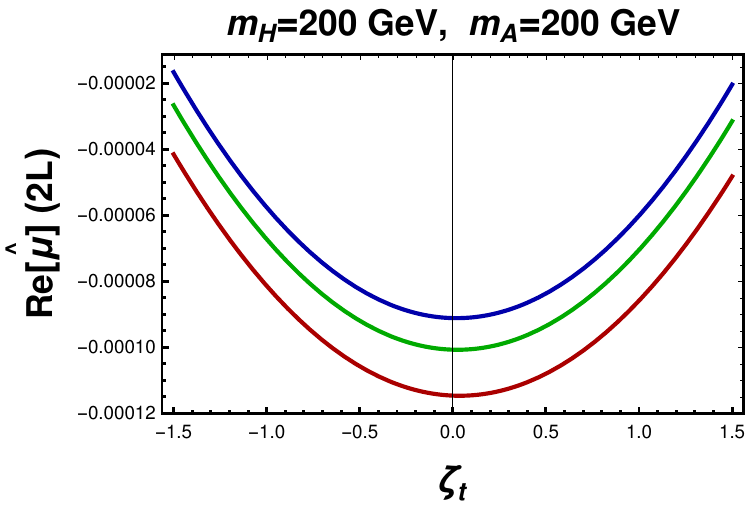}
    \includegraphics[width=0.54\linewidth]{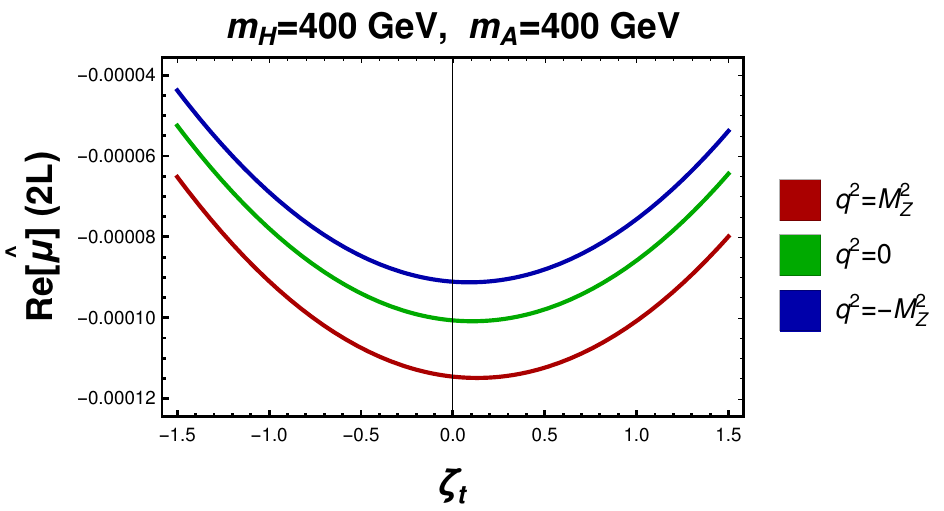}\\
    \hspace{-0.3cm}(a)\hspace{6.55cm}(b)\\
	\caption{Total two-loop contributions to the CMDM of the top quark in the ATHDM. Figures (a) and (b) show the variations of the two-loop CMDM with $\zeta_t$, where $m_{H, A}=200$~GeV and 400~GeV are assumed. Here, dark red, dark green, and dark blue colors correspond to the CMDM at $q^2=M_Z^2$, 0, and $-M_Z^2$, respectively.}
	\label{plot:twoloopATHDM}
\end{figure}

The total two-loop contributions to the top quark CMDM in the ATHDM are depicted in Fig.~\ref{plot:twoloopATHDM}a and Fig.~\ref{plot:twoloopATHDM}b for $m_{H, A}=200$~GeV and 400~GeV, respectively. The color codes are the same as in the one-loop case. In both cases, the maximum contributions to the CMDM can reach up to $\sim-1.2\times 10^{-4}$, which is one order of magnitude smaller than the total one-loop contributions (with opposite sign). It is crucial to note that the difference between the CMDMs evaluated at different scales, i.e., $q^2 = \pm M_Z^2, 0$, becomes significant in this case.  Here, the maximum value (in the negative sense) occurs at $\zeta_t\to 0$. The reason is that the CP-even and CP-odd Higgs-mediated loops interfere destructively. Also, the contributions from both diagrams are almost the same in magnitude. Now, for the CP-odd case, the Yukawa coupling is directly proportional to $\zeta_t$. Therefore, at $\zeta_t\to 0$, the total contributions come solely from the CP-even Higgs-mediated diagrams. But in the region where $|\zeta_t|\sim 1.5$, the cancellation between CP-even and CP-odd Higgs-mediated diagrams become maximum, leading to the lower value of the CMDM in those regions.

Therefore, we see that the one-loop corrections become larger in the higher $\zeta_t$ region, whereas the two-loop corrections become larger (in the negative sense) when $\zeta_t\to 0$. Since the total two-loop corrections are opposite in sign and one order of magnitude smaller compared to the total one-loop corrections, the total (one-loop $+$ two-loop) corrections become larger for larger $\zeta_t$ values.

\subsection{The CMDM of the top quark in the conventional THDMs with $\mathcal{Z}_2$ symmetries}
\label{sec:THDMz2}

Now, we discuss the one-loop and two-loop contributions of the top quark CMDM in the conventional THDMs with $\mathcal{Z}_2$ symmetries. The Yukawa couplings of the four types of THDMs are explicitly given in Eq.~\eqref{eq:convTHDMs}. Since we have no leptonic coupling involved, we have two choices: Type-I (Type-X) with $\zeta_u=\zeta_d=\frac{1}{\tan\beta}$ and Type-II (Type-Y) with $\zeta_u=-\frac{1}{\zeta_d}=\frac{1}{\tan\beta}$. Here, to calculate the CMDM, we need $\tan\beta$ and the masses of the neutral and charged Higgs. All of these parameters receive several constraints from flavor physics and theoretical considerations~\cite{Akeroyd:2016ymd, Arbey:2017gmh, Chowdhury:2017aav, Haller:2018nnx}.

In Type-I THDM, $m_H<350$~GeV is excluded for $\tan\beta\lesssim 2.5$ from the heavy Higgs searches~\cite{Chowdhury:2017aav}. Similarly, $\tan\beta\lesssim 3$ is excluded from the direct $A$ searches if $m_A<350$~GeV~\cite{Chowdhury:2017aav}. In Type-X, there is an interplay between fine-tuning and $A$ searches, which leads to a limit of $A$ to be $m_A>400$~GeV~\cite{Chowdhury:2017aav}. For Type-X, the region $m_A<400$~GeV is available only if $\tan\beta\gtrsim 10$. In type-I and type-X, $\tan\beta\lesssim 1$ region can not be accessed even if $m_H>1.5$~TeV~\cite{Chowdhury:2017aav}.

In Type-II THDM, both the scalar masses $m_H$ and $m_A$ are excluded up to $\sim 400$~GeV (for any set of values of the other THDM parameters) at 95\% CL by flavor and electroweak measurements~\cite{Haller:2018nnx}. 
From the global fit, the mass limits for the Type-II and Type-Y THDMs are $m_H>700$~GeV, $m_A>750$~GeV, $m_{H^\pm}> 740$~GeV~\cite{Chowdhury:2017aav}. Therefore, we see that there will not be any significant enhancement of the CMDM in Type-II and Type-Y. Since the mass limits are relatively smaller in the case of Type-I and Type-X THDM, we focus only on these two models.

\begin{figure}[H]
	\centering
	\includegraphics[width=0.425\linewidth]{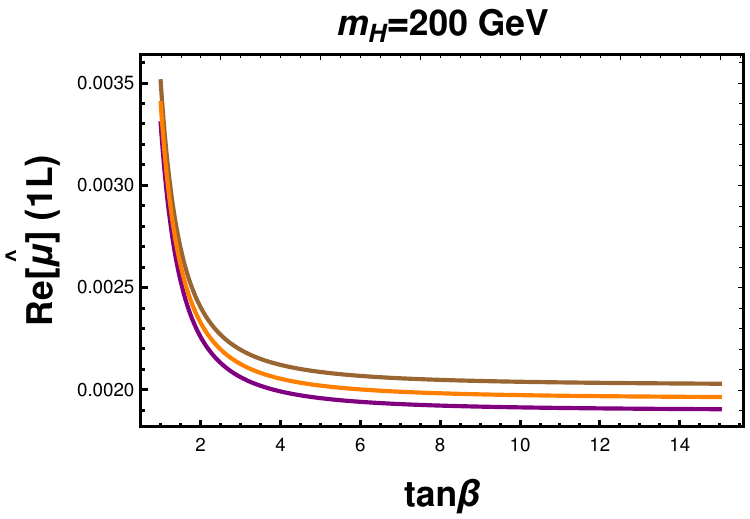}
    \includegraphics[width=0.535\linewidth]{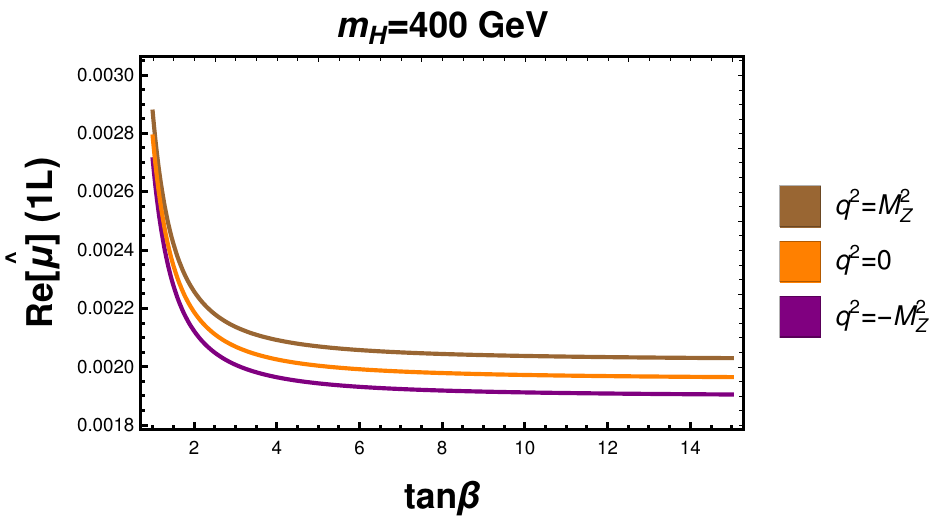}\\
    \hspace{-0.5cm}(a)\hspace{6.5cm}(b)\\
    \includegraphics[width=0.425\linewidth]{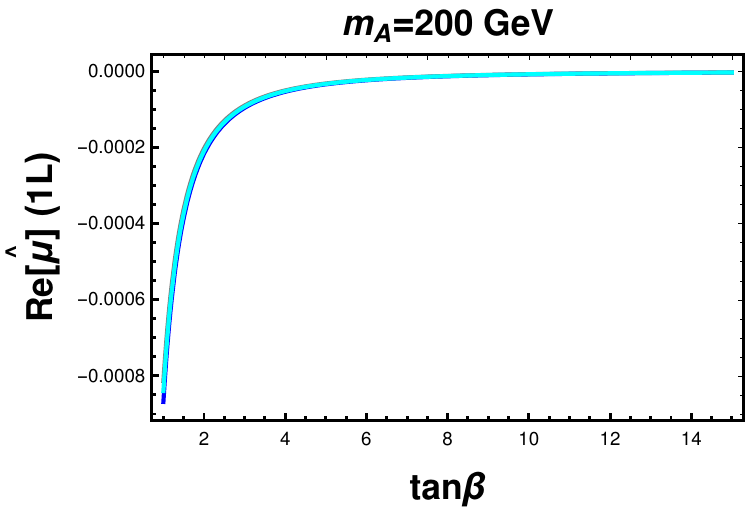}
    \includegraphics[width=0.535\linewidth]{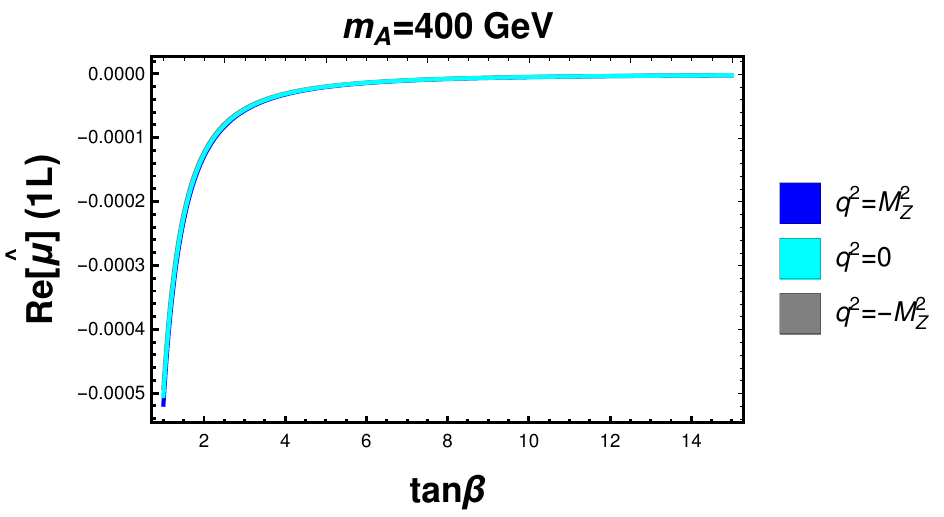}\\
    \hspace{-0.45cm}(c)\hspace{6.5cm}(d)
	\caption{One-loop contributions to the CMDM of the top quark in the Type-I (Type-X) THDM. Here, (a) and (b) show the variations of the CMDM with $\tan\beta$ for the neutral CP-even Higgs-mediated loop for $m_H=200$~GeV and 400~GeV, respectively. Figures (c) and (d) show the same for the CP-odd Higgs-mediated loop. All the color codes are the same as the previous plots.}
	\label{plot:oneloopcmdmIX}
\end{figure}

Fig.~\ref{plot:oneloopcmdmIX}a and \ref{plot:oneloopcmdmIX}b depict the variations of one-loop CMDM of the top quark with $\tan\beta$ for the CP-even Higgs-mediated diagrams where $m_H$ has been fixed at 200~GeV and 400~GeV, respectively. Here, we take the variation of $\tan\beta\in [1, 15]$. One can see that the lower values of $\tan\beta$ correspond to higher values of CMDM as $\zeta_u=\frac{1}{\tan\beta}$. The maximum value of CMDM at $q^2=M_Z^2$ for $m_H=200$~GeV can reach up to $3.5\times 10^{-3}$, whereas it reduces to $2.9\times 10^{-3}$ for $m_H=400$~GeV. Fig.~\ref{plot:oneloopcmdmIX}c and \ref{plot:oneloopcmdmIX}d show the variations of the same for the pseudoscalar Higgs-mediated diagram. The contributions from these diagrams are almost one order of magnitude smaller (with the opposite sign) than the CP-even contributions. For $m_A=200$~GeV, the maximum values of CMDM (in the negative sense) at $q^2=\pm M_Z^2$ or 0 are almost equal $\sim -8.5\times 10^{-4}$, and for $m_A=400~$GeV, the same is $\sim -5.0\times 10^{-4}$.  

\begin{figure}[H]
	\centering
	\includegraphics[width=0.425\linewidth]{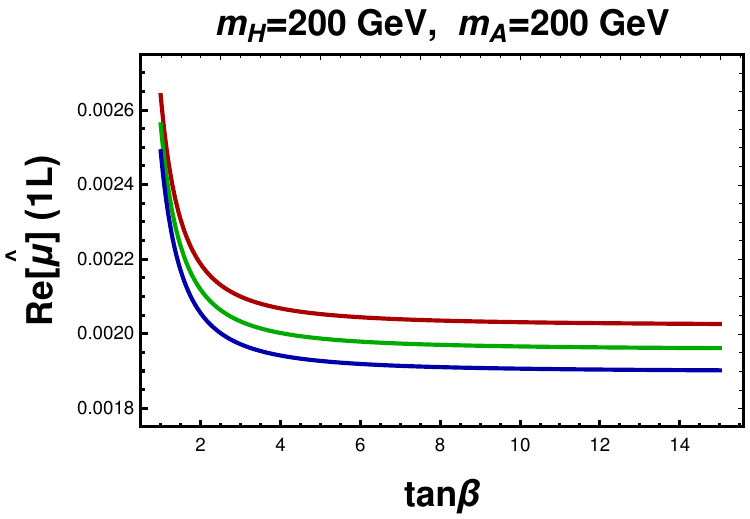}
    \includegraphics[width=0.535\linewidth]{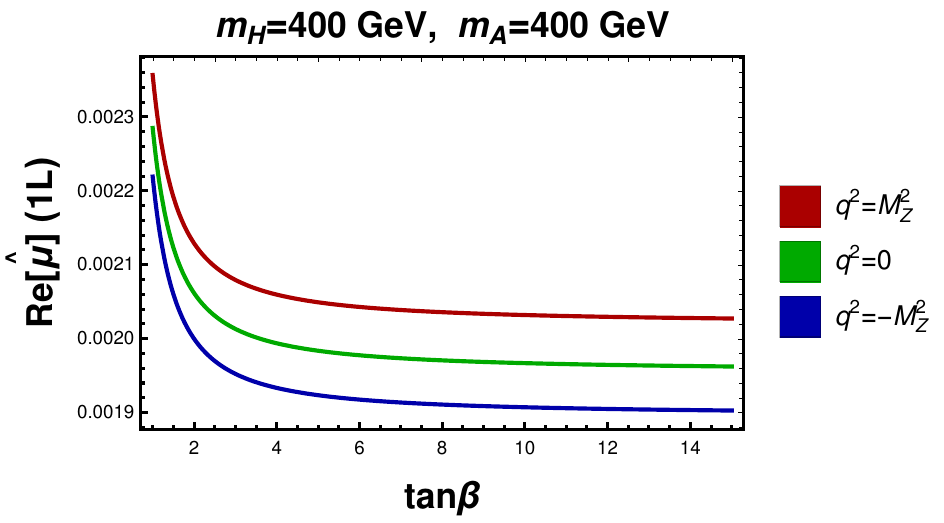}\\
    \hspace{-0.5cm}(a)\hspace{6.5cm}(b)\\
	\caption{Figures (a) and (b) show the variations of the total one-loop CMDM of the top quark with $\tan\beta$ in the Type-I (Type-X) THDM, where $m_{H, A}=200$~GeV and 400~GeV have been assumed.}
	\label{plot:oneloopTypeI}
\end{figure}

Fig.~\ref{plot:oneloopTypeI}a and \ref{plot:oneloopTypeI}b demonstrate the total one-loop corrections to the top quark CMDM in the Type-I (Type-X) THDM. The CMDM is maximum for the minimum value of $\tan\beta$. For $m_{H, A}=200$~GeV, it reaches up to $\sim 2.65\times 10^{-3}$. For an accessible parameter space point, i.e., for $\tan\beta\sim3$, Re[$\hat{\mu}$]$\simeq 1.95\times 10^{-3}$. The interesting thing is that when $m_{H, A}$ increases to 400~GeV, the overall one-loop correction turns out to be almost the same as for $m_{H, A}=200$~GeV for $\tan\beta\geq 3$. The underlying reason is that for higher values of $\tan\beta$, the heavy Higgs-mediated (both CP-even and CP-odd) diagrams have negligible contributions. Hence, for higher $\tan\beta$, the one-loop corrections are completely driven by the light Higgs, which is SM-like. So, in the higher $\tan\beta$ region, the THDM contributions become insignificant, and the CMDM value coincides with the SM contributions, which is mediated by the SM-like Higgs boson.

\begin{figure}[H]
	\centering
	\includegraphics[width=0.44\linewidth]{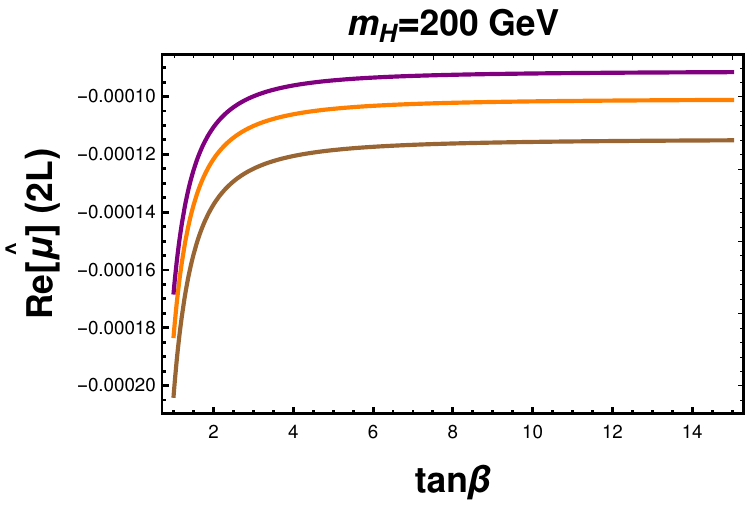}
    \includegraphics[width=0.55\linewidth]{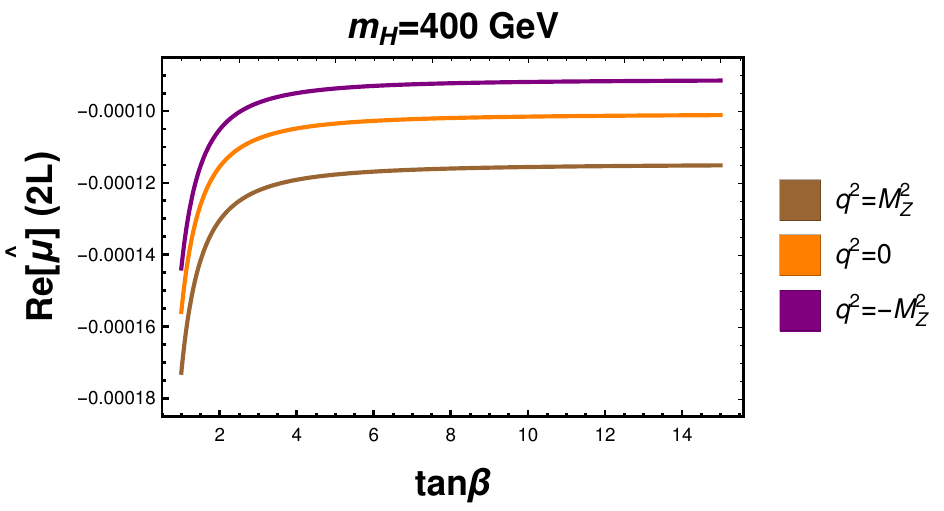}\\
    \hspace{-0.35cm}(a)\hspace{6.65cm}(b)\\
    \includegraphics[width=0.44\linewidth]{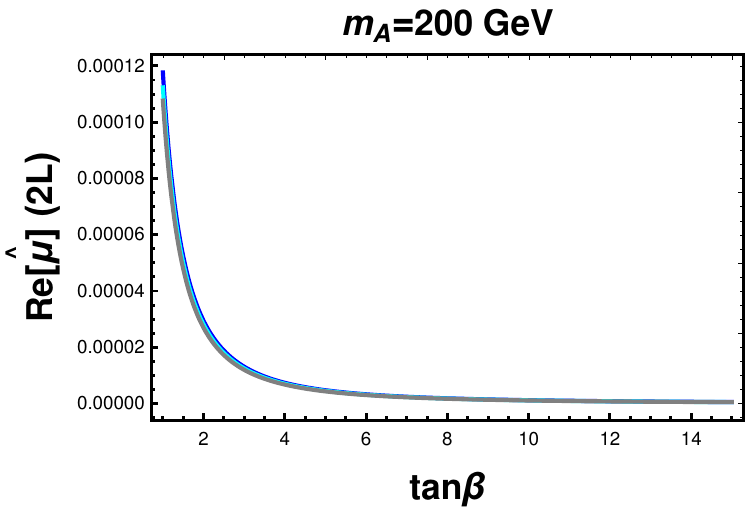}
    \includegraphics[width=0.55\linewidth]{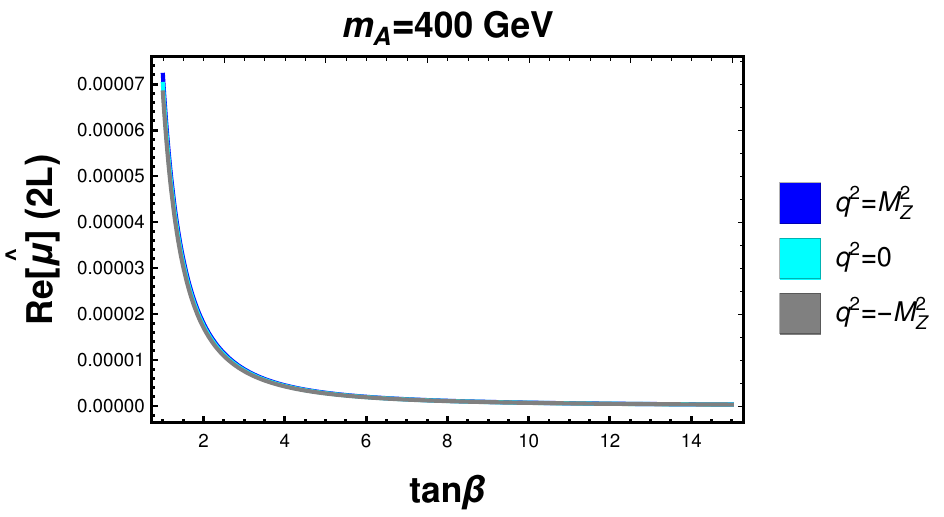}\\
    \hspace{-0.4cm}(c)\hspace{6.65cm}(d)
	\caption{Two-loop contributions to the CMDM of the top quark in the Type-I (Type-X) THDM. Figures (a) and (b) show the variations of the two-loop contributions to the CMDM with $\tan\beta$ for the CP-even Higgs scalars mediated diagrams where $m_H=200$~GeV and 400~GeV have been assumed. Figures (c) and (d) depict the same for the CP-odd scalar-mediated two-loop diagram where $m_A$ is fixed at 200~GeV and 400~GeV, respectively. The color codes are the same as the previous one.}
	\label{plot:twoloopcmdmX}
\end{figure}

The two-loop contributions to the top quark CMDM in the Type-I (Type-X) THDM are shown in Fig.~\ref{plot:twoloopcmdmX}. Fig.~\ref{plot:twoloopcmdmX}a and \ref{plot:twoloopcmdmX}b show the variations of the two-loop CMDM of the top quark with $\tan\beta$ for the CP-even Higgs-mediated diagrams where $m_H$ has been fixed at 200~GeV and 400~GeV, respectively. The maximum contribution (in the negative sense) corresponding to $\tan\beta=1$ and $m_H=200$~GeV at $q^2=M_Z^2$ is $\sim -2.05\times 10^{-4}$. For $\tan\beta\sim 3$, the contribution at $q^2=M_Z^2$ becomes $\sim -1.25\times 10^{-4}$. Similarly, Fig.~\ref{plot:twoloopcmdmX}c and \ref{plot:twoloopcmdmX}d exhibit the same for the pseudoscalar Higgs-mediated diagram. Here, the maximum contribution can reach up to $1.2\times 10^{-4}$ for $\tan\beta=1$ and $m_A=200$~GeV. For $\tan\beta\sim 3$, the contribution becomes $\sim 2\times 10^{-5}$.

\begin{figure}[H]
	\centering
	\includegraphics[width=0.44\linewidth]{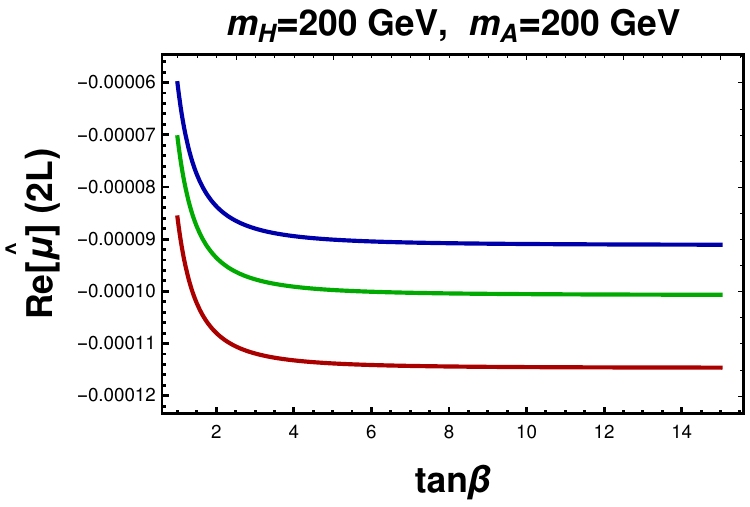}
    \includegraphics[width=0.55\linewidth]{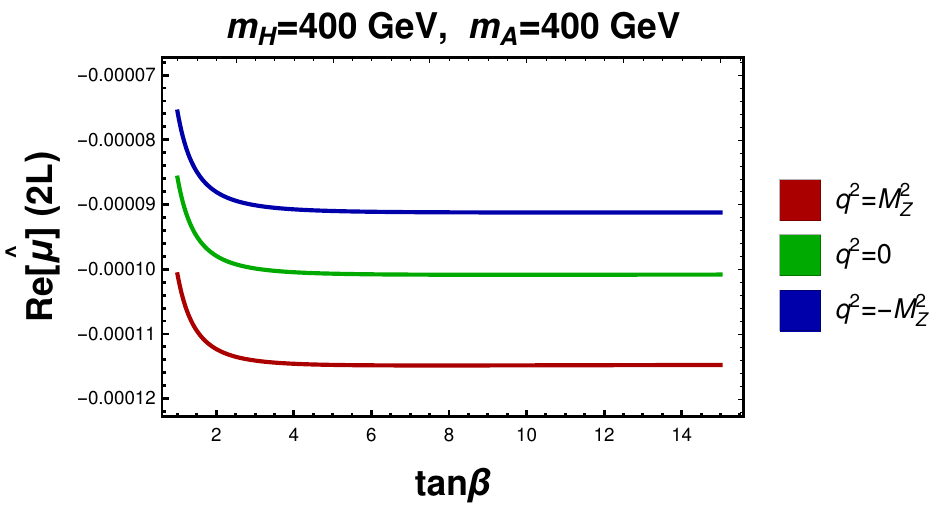}\\
    \hspace{-0.35cm}(a)\hspace{6.7cm}(b)\\
	\caption{The total two-loop contributions to the top quark CMDM  in the Type-I (Type-X) THDM. Figures (a) and (b) show the variations of total two-loop corrections to the ${\rm Re}[\hat{\mu}]$ of the top quark with $\tan\beta$ for $m_{H, A}=200$~GeV and 400~GeV, respectively. The color codes are the same as the previous one.}
	\label{plot:twoloopTypeI}
\end{figure}

The total two-loop contributions to the top quark CMDM in the Type-I (Type-X) THDM are presented in Fig.~\ref{plot:twoloopTypeI}a and \ref{plot:twoloopTypeI}b. Here, the two-loop corrections for $\tan\beta\geq 3$ are saturated to $-1.15\times 10^{-4}$ at $q^2=M_Z^2$, which is the same for both scenarios $m_{H,A}=200$~GeV and 400~GeV. For $q^2=0$ and $-M_Z^2$, the CMDM values are saturated to $-1.0\times 10^{-4}$ and $-9.0\times 10^{-5}$, respectively when $\tan\beta\geq 3$. In this case, also, for higher values of $\tan\beta$, the heavy Higgs-mediated loop corrections become insignificant.

Thus, in the conventional THDMs, we see that the one-loop corrections to the CMDM become larger when $\tan\beta\lesssim 2$ whereas the two-loop corrections become larger (in the negative sense) for larger values of $\tan\beta$, i.e., $\tan\beta>2$. Due to the cancellation effect between one-loop and two-loop corrections, the total (one-loop $+$ two-loop) corrections become larger on the lower side of $\tan\beta$. Since the lower $\tan\beta$ regions are almost excluded from the heavy Higgs searches, we do not get significant contributions to the CMDM from these conventional THDMs.

\section{Conclusions}
\label{sec:conclusion}
We have provided analytical calculations for the one-loop and two-loop contributions to the CMDM of quarks, considering the off-shell nature of the external gluon.
Throughout the analytical calculation, we have not imposed any approximations and have presented the analytical results in a model-independent manner. 
Then, we numerically evaluated the CMDM of the top quark in both the ATHDM and the conventional THDMs with $\mathcal{Z}_2$ symmetries, considering the momentum transfer of the external gluon, $q^2=\pm M_Z^2$ and $0$. The THDMs we have considered here offer contributions of $\mathcal{O}(10^{-3})$ at one-loop and $\mathcal{O}(10^{-4})$ at two-loop level. The interference between the total contributions from the one-loop and two-loop levels is destructive in all types of THDMs considered, resulting in a total contribution (one-loop $+$ two-loop) on the order of $\mathcal{O}(10^{-3})$ with a positive sign.  
Since these are the only contributions to the CMDM up to the two-loop level in the THDMs, these contributions may be added to the SM contributions, leading to more precise theoretical predictions from the BSM side.
The analytical results presented here are completely general and can be used in any BSM model with extended Higgs sectors.

\section{ACKNOWLEDGEMENTS}
The computations were supported in part by SAMKHYA, the High-Performance Computing Facility provided by the Institute of Physics (IoP), Bhubaneswar, India. I would like to acknowledge D. Das for his valuable discussions.
\vfill
\bibliographystyle{JHEPCust.bst}
\bibliography{CMDM}

\providecommand{\href}[2]{#2}\begingroup\raggedright\begin{thebibliography}{10}

\bibitem{CMS:2019kzp}
{\scshape CMS} collaboration, A.~M. Sirunyan et~al., \emph{{Measurement of the
  top quark forward-backward production asymmetry and the anomalous
  chromoelectric and chromomagnetic moments in pp collisions at $ \sqrt{s} $ =
  13 TeV}}, \href{http://dx.doi.org/10.1007/JHEP06(2020)146}{\emph{JHEP} {\bf
  06} (2020) 146}, [\href{http://arxiv.org/abs/1912.09540}{{\tt 1912.09540}}].

\bibitem{Choudhury:2014lna}
I.~D. Choudhury and A.~Lahiri, \emph{{Anomalous chromomagnetic moment of
  quarks}}, \href{http://dx.doi.org/10.1142/S0217732315501138}{\emph{Mod. Phys.
  Lett. A} {\bf 30} (2015) 1550113},
  [\href{http://arxiv.org/abs/1409.0073}{{\tt 1409.0073}}].

\bibitem{Aranda:2018zis}
J.~I. Aranda, D.~Espinosa-G\'omez, J.~Monta\~no, B.~Quezadas-Vivian,
  F.~Ram\'\i{}rez-Zavaleta and E.~S. Tututi, \emph{{Flavor violation in chromo-
  and electromagnetic dipole moments induced by Z' gauge bosons and a brief
  revisit of the Standard Model}},
  \href{http://dx.doi.org/10.1103/PhysRevD.98.116003}{\emph{Phys. Rev. D} {\bf
  98} (2018) 116003}, [\href{http://arxiv.org/abs/1809.02817}{{\tt
  1809.02817}}].

\bibitem{Aranda:2020tox}
J.~I. Aranda, T.~Cisneros-P\'erez, J.~Monta\~no, B.~Quezadas-Vivian,
  F.~Ram\'\i{}rez-Zavaleta and E.~S. Tututi, \emph{{Revisiting the top quark
  chromomagnetic dipole moment in the SM}},
  \href{http://dx.doi.org/10.1140/epjp/s13360-021-01102-x}{\emph{Eur. Phys. J.
  Plus} {\bf 136} (2021) 164}, [\href{http://arxiv.org/abs/2009.05195}{{\tt
  2009.05195}}].

\bibitem{Hernandez-Juarez:2020drn}
A.~I. Hern\'andez-Ju\'arez, A.~Moyotl and G.~Tavares-Velasco, \emph{{New
  estimate of the chromomagnetic dipole moment of quarks in the standard
  model}}, \href{http://dx.doi.org/10.1140/epjp/s13360-021-01239-9}{\emph{Eur.
  Phys. J. Plus} {\bf 136} (2021) 262},
  [\href{http://arxiv.org/abs/2009.11955}{{\tt 2009.11955}}].

\bibitem{Bermudez:2017bpx}
R.~Bermudez, L.~Albino, L.~X. Guti\'errez-Guerrero, M.~E. Tejeda-Yeomans and
  A.~Bashir, \emph{{Quark-gluon Vertex: A Perturbation Theory Primer and
  Beyond}}, \href{http://dx.doi.org/10.1103/PhysRevD.95.034041}{\emph{Phys.
  Rev. D} {\bf 95} (2017) 034041}, [\href{http://arxiv.org/abs/1702.04437}{{\tt
  1702.04437}}].

\bibitem{ParticleDataGroup:2020ssz}
{\scshape Particle Data Group} collaboration, P.~A. Zyla et~al., \emph{{Review
  of Particle Physics}},
  \href{http://dx.doi.org/10.1093/ptep/ptaa104}{\emph{PTEP} {\bf 2020} (2020)
  083C01}.

\bibitem{Gaitan:2015aia}
R.~Gaitan, E.~A. Garces, J.~H.~M. de~Oca and R.~Martinez, \emph{{Top quark
  Chromoelectric and Chromomagnetic Dipole Moments in a Two Higgs Doublet Model
  with CP violation}},
  \href{http://dx.doi.org/10.1103/PhysRevD.92.094025}{\emph{Phys. Rev. D} {\bf
  92} (2015) 094025}, [\href{http://arxiv.org/abs/1505.04168}{{\tt
  1505.04168}}].

\bibitem{Hernandez-Juarez:2018uow}
A.~I. Hern\'andez-Ju\'arez, A.~Moyotl and G.~Tavares-Velasco,
  \emph{{Chromomagnetic and chromoelectric dipole moments of the top quark in
  the fourth-generation THDM}},
  \href{http://dx.doi.org/10.1103/PhysRevD.98.035040}{\emph{Phys. Rev. D} {\bf
  98} (2018) 035040}, [\href{http://arxiv.org/abs/1805.00615}{{\tt
  1805.00615}}].

\bibitem{Martinez:2001qs}
R.~Martinez and J.~A. Rodriguez, \emph{{The Anomalous chromomagnetic dipole
  moment of the top quark in the standard model and beyond}},
  \href{http://dx.doi.org/10.1103/PhysRevD.65.057301}{\emph{Phys. Rev. D} {\bf
  65} (2002) 057301}, [\href{http://arxiv.org/abs/hep-ph/0109109}{{\tt
  hep-ph/0109109}}].

\bibitem{Martinez:2007qf}
R.~Martinez, M.~A. Perez and N.~Poveda, \emph{{Chromomagnetic Dipole Moment of
  the Top Quark Revisited}},
  \href{http://dx.doi.org/10.1140/epjc/s10052-007-0457-6}{\emph{Eur. Phys. J.
  C} {\bf 53} (2008) 221--230},
  [\href{http://arxiv.org/abs/hep-ph/0701098}{{\tt hep-ph/0701098}}].

\bibitem{Cao:2008qd}
Q.-H. Cao, C.-R. Chen, F.~Larios and C.~P. Yuan, \emph{{Anomalous gtt couplings
  in the Littlest Higgs Model with T-parity}},
  \href{http://dx.doi.org/10.1103/PhysRevD.79.015004}{\emph{Phys. Rev. D} {\bf
  79} (2009) 015004}, [\href{http://arxiv.org/abs/0801.2998}{{\tt 0801.2998}}].

\bibitem{Ding:2008nh}
L.~Ding and C.-X. Yue, \emph{{Top quark chromomagnetic dipole moment in the
  littlest Higgs model with T-parity}},
  \href{http://dx.doi.org/10.1088/0253-6102/50/2/32}{\emph{Commun. Theor.
  Phys.} {\bf 50} (2008) 441--444}, [\href{http://arxiv.org/abs/0801.1880}{{\tt
  0801.1880}}].

\bibitem{Cisneros-Perez:2024onx}
T.~Cisneros-P\'erez, M.~A. Hern\'andez-Ru\'\i{}z, A.~Ramirez-Morales,
  A.~Guti\'errez-Rodr\'\i{}guez and J.~Monta\~no Dom\'\i{}nguez, \emph{{Top
  anomalous chromomagnetic dipole moment in the Bestest Little Higgs Model}},
  \href{http://arxiv.org/abs/2403.08021}{{\tt 2403.08021}}.

\bibitem{Aboubrahim:2015zpa}
A.~Aboubrahim, T.~Ibrahim, P.~Nath and A.~Zorik, \emph{{Chromoelectric Dipole
  Moments of Quarks in MSSM Extensions}},
  \href{http://dx.doi.org/10.1103/PhysRevD.92.035013}{\emph{Phys. Rev. D} {\bf
  92} (2015) 035013}, [\href{http://arxiv.org/abs/1507.02668}{{\tt
  1507.02668}}].

\bibitem{Martinez:2008hm}
R.~Martinez, M.~A. Perez and O.~A. Sampayo, \emph{{Constraints on unparticle
  physics from the gt $\bar{t}$ anomalous coupling}},
  \href{http://dx.doi.org/10.1142/S0217751X10048159}{\emph{Int. J. Mod. Phys.
  A} {\bf 25} (2010) 1061--1067}, [\href{http://arxiv.org/abs/0805.0371}{{\tt
  0805.0371}}].

\bibitem{Ibrahim:2011im}
T.~Ibrahim and P.~Nath, \emph{{The Chromoelectric Dipole Moment of the Top
  Quark in Models with Vector Like Multiplets}},
  \href{http://dx.doi.org/10.1103/PhysRevD.84.015003}{\emph{Phys. Rev. D} {\bf
  84} (2011) 015003}, [\href{http://arxiv.org/abs/1104.3851}{{\tt 1104.3851}}].

\bibitem{Hernandez-Juarez:2020gxp}
A.~I. Hern\'andez-Ju\'arez, G.~Tavares-Velasco and A.~Moyotl,
  \emph{{Chromomagnetic and chromoelectric dipole moments of quarks in the
  reduced 331 model}},
  \href{http://dx.doi.org/10.1088/1674-1137/ac1b9a}{\emph{Chin. Phys. C} {\bf
  45} (2021) 113101}, [\href{http://arxiv.org/abs/2012.09883}{{\tt
  2012.09883}}].

\bibitem{Martinez:1996cy}
R.~Martinez and J.~A. Rodriguez, \emph{{Using the radiative decay $b \to s
  \gamma$ to bound the chromomagnetic dipole moment of the top quark}},
  \href{http://dx.doi.org/10.1103/PhysRevD.55.3212}{\emph{Phys. Rev. D} {\bf
  55} (1997) 3212--3214}, [\href{http://arxiv.org/abs/hep-ph/9612438}{{\tt
  hep-ph/9612438}}].

\bibitem{BuarqueFranzosi:2015jrv}
D.~Buarque~Franzosi and C.~Zhang, \emph{{Probing the top-quark chromomagnetic
  dipole moment at next-to-leading order in QCD}},
  \href{http://dx.doi.org/10.1103/PhysRevD.91.114010}{\emph{Phys. Rev. D} {\bf
  91} (2015) 114010}, [\href{http://arxiv.org/abs/1503.08841}{{\tt
  1503.08841}}].

\bibitem{Pich:2009sp}
A.~Pich and P.~Tuzon, \emph{{Yukawa Alignment in the Two-Higgs-Doublet Model}},
  \href{http://dx.doi.org/10.1103/PhysRevD.80.091702}{\emph{Phys. Rev. D} {\bf
  80} (2009) 091702}, [\href{http://arxiv.org/abs/0908.1554}{{\tt 0908.1554}}].

\bibitem{Pich:2010ic}
A.~Pich, \emph{{Flavour constraints on multi-Higgs-doublet models: Yukawa
  alignment}},
  \href{http://dx.doi.org/10.1016/j.nuclphysbps.2010.12.030}{\emph{Nucl. Phys.
  B Proc. Suppl.} {\bf 209} (2010) 182--187},
  [\href{http://arxiv.org/abs/1010.5217}{{\tt 1010.5217}}].

\bibitem{Karan:2023kyj}
A.~Karan, V.~Miralles and A.~Pich, \emph{{Updated global fit of the aligned
  two-Higgs-doublet model with heavy scalars}},
  \href{http://dx.doi.org/10.1103/PhysRevD.109.035012}{\emph{Phys. Rev. D} {\bf
  109} (2024) 035012}, [\href{http://arxiv.org/abs/2307.15419}{{\tt
  2307.15419}}].

\bibitem{Eberhardt:2020dat}
O.~Eberhardt, A.~P.~n. Mart\'\i{}nez and A.~Pich, \emph{{Global fits in the
  Aligned Two-Higgs-Doublet model}},
  \href{http://dx.doi.org/10.1007/JHEP05(2021)005}{\emph{JHEP} {\bf 05} (2021)
  005}, [\href{http://arxiv.org/abs/2012.09200}{{\tt 2012.09200}}].

\bibitem{Eberhardt:2013uba}
O.~Eberhardt, U.~Nierste and M.~Wiebusch, \emph{{Status of the
  two-Higgs-doublet model of type II}},
  \href{http://dx.doi.org/10.1007/JHEP07(2013)118}{\emph{JHEP} {\bf 07} (2013)
  118}, [\href{http://arxiv.org/abs/1305.1649}{{\tt 1305.1649}}].

\bibitem{Chowdhury:2015yja}
D.~Chowdhury and O.~Eberhardt, \emph{{Global fits of the two-loop renormalized
  Two-Higgs-Doublet model with soft Z$_{2}$ breaking}},
  \href{http://dx.doi.org/10.1007/JHEP11(2015)052}{\emph{JHEP} {\bf 11} (2015)
  052}, [\href{http://arxiv.org/abs/1503.08216}{{\tt 1503.08216}}].

\bibitem{Cacchio:2016qyh}
V.~Cacchio, D.~Chowdhury, O.~Eberhardt and C.~W. Murphy, \emph{{Next-to-leading
  order unitarity fits in Two-Higgs-Doublet models with soft $\mathbb{Z}_2$
  breaking}}, \href{http://dx.doi.org/10.1007/JHEP11(2016)026}{\emph{JHEP} {\bf
  11} (2016) 026}, [\href{http://arxiv.org/abs/1609.01290}{{\tt 1609.01290}}].

\bibitem{Eberhardt:2018lub}
O.~Eberhardt, \emph{{Current status of Two-Higgs-Doublet models with a softly
  broken $\mathbb{Z}_2$ symmetry}},
  \href{http://dx.doi.org/10.22323/1.340.0457}{\emph{PoS} {\bf ICHEP2018}
  (2019) 457}, [\href{http://arxiv.org/abs/1809.04851}{{\tt 1809.04851}}].

\bibitem{Chowdhury:2017aav}
D.~Chowdhury and O.~Eberhardt, \emph{{Update of Global Two-Higgs-Doublet Model
  Fits}}, \href{http://dx.doi.org/10.1007/JHEP05(2018)161}{\emph{JHEP} {\bf 05}
  (2018) 161}, [\href{http://arxiv.org/abs/1711.02095}{{\tt 1711.02095}}].

\bibitem{Eberhardt:2017ulj}
O.~Eberhardt, \emph{{Two-Higgs-doublet model fits with HEPfit}},  in
  \emph{{2017 European Physical Society Conference on High Energy Physics}}, 9,
  2017.
\newblock \href{http://arxiv.org/abs/1709.09414}{{\tt 1709.09414}}.
\newblock \href{http://dx.doi.org/10.22323/1.314.0281}{DOI}.

\bibitem{Eberhardt:2015ypa}
O.~Eberhardt, \emph{{Fitting the Two-Loop Renormalized Two-Higgs-Doublet
  Model}}, {\emph{PoS} {\bf PLANCK2015} (2015) 040},
  [\href{http://arxiv.org/abs/1510.05966}{{\tt 1510.05966}}].

\bibitem{Eberhardt:2014kaa}
O.~Eberhardt, \emph{{Fitting the Two-Higgs-Doublet model of type II}},  in
  \emph{{49th Rencontres de Moriond on Electroweak Interactions and Unified
  Theories}}, pp.~523--526, 2014.
\newblock \href{http://arxiv.org/abs/1405.3181}{{\tt 1405.3181}}.

\bibitem{Gunion:2002zf}
J.~F. Gunion and H.~E. Haber, \emph{{The CP conserving two Higgs doublet model:
  The Approach to the decoupling limit}},
  \href{http://dx.doi.org/10.1103/PhysRevD.67.075019}{\emph{Phys. Rev. D} {\bf
  67} (2003) 075019}, [\href{http://arxiv.org/abs/hep-ph/0207010}{{\tt
  hep-ph/0207010}}].

\bibitem{Barr:1990vd}
S.~M. Barr and A.~Zee, \emph{{Electric Dipole Moment of the Electron and of the
  Neutron}}, \href{http://dx.doi.org/10.1103/PhysRevLett.65.21}{\emph{Phys.
  Rev. Lett.} {\bf 65} (1990) 21--24}. [Erratum: Phys.Rev.Lett. 65, 2920
  (1990)].

\bibitem{Bernreuther:2013aga}
W.~Bernreuther and Z.-G. Si, \emph{{Top quark spin correlations and
  polarization at the LHC: standard model predictions and effects of anomalous
  top chromo moments}},
  \href{http://dx.doi.org/10.1016/j.physletb.2013.06.051}{\emph{Phys. Lett. B}
  {\bf 725} (2013) 115--122}, [\href{http://arxiv.org/abs/1305.2066}{{\tt
  1305.2066}}]. [Erratum: Phys.Lett.B 744, 413--413 (2015)].

\bibitem{CMS:2016piu}
{\scshape CMS} collaboration, V.~Khachatryan et~al., \emph{{Measurements of t
  t-bar spin correlations and top quark polarization using dilepton final
  states in pp collisions at sqrt(s) = 8 TeV}},
  \href{http://dx.doi.org/10.1103/PhysRevD.93.052007}{\emph{Phys. Rev. D} {\bf
  93} (2016) 052007}, [\href{http://arxiv.org/abs/1601.01107}{{\tt
  1601.01107}}].

\bibitem{Haberl:1995ek}
P.~Haberl, O.~Nachtmann and A.~Wilch, \emph{{Top production in hadron hadron
  collisions and anomalous top - gluon couplings}},
  \href{http://dx.doi.org/10.1103/PhysRevD.53.4875}{\emph{Phys. Rev. D} {\bf
  53} (1996) 4875--4885}, [\href{http://arxiv.org/abs/hep-ph/9505409}{{\tt
  hep-ph/9505409}}].

\bibitem{Czarnecki:1997bu}
A.~Czarnecki and B.~Krause, \emph{{Neutron electric dipole moment in the
  standard model: Valence quark contributions}},
  \href{http://dx.doi.org/10.1103/PhysRevLett.78.4339}{\emph{Phys. Rev. Lett.}
  {\bf 78} (1997) 4339--4342}, [\href{http://arxiv.org/abs/hep-ph/9704355}{{\tt
  hep-ph/9704355}}].

\bibitem{Patel:2015tea}
H.~H. Patel, \emph{{Package-X: A Mathematica package for the analytic
  calculation of one-loop integrals}},
  \href{http://dx.doi.org/10.1016/j.cpc.2015.08.017}{\emph{Comput. Phys.
  Commun.} {\bf 197} (2015) 276--290},
  [\href{http://arxiv.org/abs/1503.01469}{{\tt 1503.01469}}].

\bibitem{Patel:2016fam}
H.~H. Patel, \emph{{Package-X 2.0: A Mathematica package for the analytic
  calculation of one-loop integrals}},
  \href{http://dx.doi.org/10.1016/j.cpc.2017.04.015}{\emph{Comput. Phys.
  Commun.} {\bf 218} (2017) 66--70},
  [\href{http://arxiv.org/abs/1612.00009}{{\tt 1612.00009}}].

\bibitem{Bisal:2022nbn}
S.~Bisal, D.~Das, S.~Majhi and S.~Mitra, \emph{{Production of singlet dominated
  scalar(s) at the LHC}},
  \href{http://dx.doi.org/10.1016/j.physletb.2023.137806}{\emph{Phys. Lett. B}
  {\bf 839} (2023) 137806}, [\href{http://arxiv.org/abs/2207.01358}{{\tt
  2207.01358}}].

\bibitem{Celis:2013rcs}
A.~Celis, V.~Ilisie and A.~Pich, \emph{{LHC constraints on two-Higgs doublet
  models}}, \href{http://dx.doi.org/10.1007/JHEP07(2013)053}{\emph{JHEP} {\bf
  07} (2013) 053}, [\href{http://arxiv.org/abs/1302.4022}{{\tt 1302.4022}}].

\bibitem{Manohar:2006ga}
A.~V. Manohar and M.~B. Wise, \emph{{Flavor changing neutral currents, an
  extended scalar sector, and the Higgs production rate at the CERN LHC}},
  \href{http://dx.doi.org/10.1103/PhysRevD.74.035009}{\emph{Phys. Rev. D} {\bf
  74} (2006) 035009}, [\href{http://arxiv.org/abs/hep-ph/0606172}{{\tt
  hep-ph/0606172}}].

\bibitem{Allwicher:2021rtd}
L.~Allwicher, P.~Arnan, D.~Barducci and M.~Nardecchia, \emph{{Perturbative
  unitarity constraints on generic Yukawa interactions}},
  \href{http://dx.doi.org/10.1007/JHEP10(2021)129}{\emph{JHEP} {\bf 10} (2021)
  129}, [\href{http://arxiv.org/abs/2108.00013}{{\tt 2108.00013}}].

\bibitem{Jung:2010ik}
M.~Jung, A.~Pich and P.~Tuzon, \emph{{Charged-Higgs phenomenology in the
  Aligned two-Higgs-doublet model}},
  \href{http://dx.doi.org/10.1007/JHEP11(2010)003}{\emph{JHEP} {\bf 11} (2010)
  003}, [\href{http://arxiv.org/abs/1006.0470}{{\tt 1006.0470}}].

\bibitem{Akeroyd:2016ymd}
A.~G. Akeroyd et~al., \emph{{Prospects for charged Higgs searches at the LHC}},
  \href{http://dx.doi.org/10.1140/epjc/s10052-017-4829-2}{\emph{Eur. Phys. J.
  C} {\bf 77} (2017) 276}, [\href{http://arxiv.org/abs/1607.01320}{{\tt
  1607.01320}}].

\bibitem{Arbey:2017gmh}
A.~Arbey, F.~Mahmoudi, O.~Stal and T.~Stefaniak, \emph{{Status of the Charged
  Higgs Boson in Two Higgs Doublet Models}},
  \href{http://dx.doi.org/10.1140/epjc/s10052-018-5651-1}{\emph{Eur. Phys. J.
  C} {\bf 78} (2018) 182}, [\href{http://arxiv.org/abs/1706.07414}{{\tt
  1706.07414}}].

\bibitem{Haller:2018nnx}
J.~Haller, A.~Hoecker, R.~Kogler, K.~M\"onig, T.~Peiffer and J.~Stelzer,
  \emph{{Update of the global electroweak fit and constraints on
  two-Higgs-doublet models}},
  \href{http://dx.doi.org/10.1140/epjc/s10052-018-6131-3}{\emph{Eur. Phys. J.
  C} {\bf 78} (2018) 675}, [\href{http://arxiv.org/abs/1803.01853}{{\tt
  1803.01853}}].

\end{thebibliography}\endgroup
\vfill
\end{document}